\documentclass[11pt,a4paper,english,oneside]{article}

\usepackage{authblk} 

\usepackage[utf8]{inputenc}
\usepackage{newunicodechar}
\newunicodechar{−}{\textminus}

\usepackage[left=2.5cm,top=2.5cm,right=2.5cm,bottom=2cm]{geometry}

\usepackage{amsmath, amssymb, amsthm}

\usepackage{makeidx}
\makeindex

\usepackage{fancyhdr}
\usepackage{graphicx}
\usepackage{subcaption}
\usepackage{epsfig}

\usepackage{multirow}
\usepackage{cancel}
\usepackage{adjustbox}
\usepackage{float}
\usepackage{caption}
\usepackage{hyperref}
\usepackage{quiver}
\usepackage{enumitem}
\usepackage[most]{tcolorbox}
\usepackage{xcolor}
\usepackage{setspace}
\usepackage{parskip}
\usepackage{afterpage}
\usepackage{csquotes}
\usepackage[backend=biber, sorting=none, style=numeric, maxbibnames=99]{biblatex}
\def \punt {\leaders \hbox to 4mm {\hfil . \hfil } \hfill}

\theoremstyle{definition}

\newcommand{\R}{\mathbb R}

\newcommand{\N}{\mathbb N}

\makeatletter
\newcommand{\skipitems}[1]{%
  \addtocounter{\@enumctr}{#1}%
}
\makeatother

\title{Can PCE solve the factorisation problem via optimisation?}

\author[1,2]{Fernando Alonso}
\author[1]{Colomán Samprón}
\author[1]{Jacobo Veiga}
\author[1,*]{Andrés Gómez}

\affil[1]{Galicia Supercomputing Center (CESGA), Spain}
\affil[2]{Centro de Investigación TIC (CITIC), Spain}
\affil[*]{Corresponding author: agomez@cesga.es}
\date{\today}

\begin{document}

\maketitle

\begin{abstract}
The ongoing progress in quantum technologies has fueled a sustained exploration of their potential applications across various domains, particularly in computational problems that are considered intractable for classical systems. Among these problems, integer factorisation remains of special interest due to its relevance to widely used cryptographic schemes such as RSA. Among the different possibilities, one approach to factorisation is to convert the problem into a binary optimisation problem. However, current proposals usually need a large number of qubits that make them unfeasible within the current hardware. In this work, we investigate a possible adaptation of the Pauli Correlation Encoding (PCE) algorithm to the factorisation problem. Due to its compression capability, it can drastically reduce the number of needed qubits. The proposed approach explores how the structure and dynamics of the PCE framework may be employed to encode and analyze candidate factor relations within a quantum computational setting. Rather than presenting a replacement for established quantum factorisation methods, this study aims to provide a preliminary examination of the feasibility and limitations of the proposed adaptation. We discuss the algorithmic design, its conceptual relationship with existing quantum approaches, and the practical constraints associated with implementation on current or near-term quantum hardware. Initial observations suggest that the method may offer an alternative perspective for studying factorisation within the broader context of quantum computation, although no claim is made regarding computational advantage. These results are intended primarily as an exploratory contribution to ongoing research in quantum algorithms and computational number theory.

\end{abstract}



\section{Introduction}

Number theory has been a highly prolific field of study throughout history, primarily devoted to the study of integers and arithmetic, as well as to the investigation of their properties and characteristics, among them primality. The motivation for studying the properties of prime numbers arises not only from the theoretical understanding of mathematical foundations, but also from their numerous applications. More specifically, since the formulation and proof of the Fundamental Theorem of Arithmetic, which states that every positive integer higher than 1 is either prime or can be expressed uniquely as a finite product of prime numbers, one of the main areas of interest has been the general resolution of the integer factorisation problem, as prime numbers constitute the fundamental basis of integer factor decomposition.

Thus, given $\mathbb{P} \subset \mathbb{N}$ as the set of prime numbers, the general integer factorisation problem consists of determining the prime factor decomposition of a composite integer $N \in \mathbb{N}$. Mathematically, it is based on finding a set of prime numbers
\begin{equation*}
    \Big\{1<p_i<N ~\big|~i\in\{1,\ldots,s\}\Big\}
\end{equation*}
and a number of indexes 
\begin{equation*}
    \Big\{\alpha_i \in \mathbb{N}~\big|~~i\in\{1,\ldots,s\}\Big\},
\end{equation*}
such that
\begin{equation*}
N = \prod_{i=1}^{s} p_i^{\alpha_i}.
\end{equation*}
In the particular case where the composite integer $N \in \mathbb{N}$ can be expressed as the product of exactly two prime numbers, that is, if there exist $p,q \in \mathbb{P}$ such that $N = p \cdot q$, we say that $N$ is a semiprime number.

The main interest in solving the factorisation problem for semiprime numbers lies in its computational difficulty, since the RSA public-key cryptographic system, one of the most widely known and currently used cryptographic schemes, is based upon it \cite{Gidney_2025}. RSA enables communicating parties to exchange messages securely through an insecure channel without the need to previously share a private key, as the robustness of the encryption relies on the difficulty of finding the value $x$ capable of solving equations of the form
\begin{equation}
x^e \equiv c \pmod{N},
\end{equation}
when $e$, $c$, and $N$ are known numbers, and furthermore $N$ is the product of two large prime numbers.


Both the integer factorisation problem and the semiprime factorisation problem constitute some of the most challenging problems in computation, since, to date, the existence of a classical algorithm capable of solving integer factorisation in polynomial time has not been demonstrated. Furthermore, from the perspective of computational complexity theory, this problem belongs to the NP class, as a prime factor decomposition can be efficiently verified; however, it has neither been proven to belong to the class P nor to be NP-complete.

Despite this, numerous factorisation algorithms have been proposed \cite{Bansimba_2025}. Some are specialised-purpose methods, such as \textit{Fermat’s Method}~\cite{Gupta_2009}, \textit{Euler’s Method}~\cite{Alberts_2026}, or the \textit{Special Number Field Sieve}~\cite{Pomerance_1998}; while others are general-purpose algorithms, such as \textit{Dixon’s Method}~\cite{Lenstra_2005}, the \textit{Quadratic Sieve}~\cite{Pomerance_1984}, or the \textit{General Number Field Sieve}~\cite{Lenstra_2001}. The latter is currently regarded as the most efficient known method for factoring integers. In fact, this algorithm enabled the factorisation of the largest known RSA number to date, RSA-250, a 250-decimal-digit integer (829 bits), through the use of thousands of accumulated computation years on distributed high-performance computing architectures.

All these hardware resource requirements and the need for extremely high computational power highlight the importance of exploring new directions and proposing novel approaches that can contribute to the efficient resolution of such a relevant problem in cryptography and number theory.

In this context, an important development was introduced in 1994 by Peter W. Shor, who proposed a quantum algorithm for semiprime integer factorisation \footnote{From this point onward, unless explicitly stated otherwise, the terms \textit{factorization problem}  and \textit{integer factorization} will refer to the \textit{semiprime factorization problem} and  \textit{semiprime integer factorization}.} with polynomial-time complexity on a quantum computer~\cite{Shor_1997}. This result established that quantum computation could efficiently solve the factorisation problem, in contrast to the best-known classical algorithms, and highlighted the potential implications of quantum technologies for cryptographic systems based on the hardness of factoring large integers. Since then, substantial efforts have been devoted to the development of both quantum and quantum-inspired algorithms, as well as to the advancement of quantum hardware. Motivated by the possibility of exploiting quantum phenomena and quantum-inspired techniques to address computationally challenging problems, these developments have expanded the range of quantum computational approaches and their potential applications.

However, despite its theoretical significance, the practical implementation of Shor’s algorithm remains beyond the capabilities of current \textit{Noisy Intermediate-Scale Quantum} (NISQ) devices~\cite{Preskill_2018}. In this era, quantum computers are typically limited to systems comprising tens or hundreds of qubits, whose performance is further constrained by noise, decoherence, and imperfect gate operations. In particular, the number of logical qubits and the overall quantum resources required to factor integers of cryptographic relevance, such as those employed in RSA encryption, remain prohibitively large, rendering the practical execution of Shor’s algorithm infeasible with present-day hardware~\cite{Bagourd_2025}.

Nevertheless, important experimental and computational milestones have been achieved in the implementation of Shor’s algorithm. To the best of our knowledge, one of the largest experimental demonstrations on a real quantum computer is reported in~\cite{Amico_2019}, where the authors implemented a compiled version of Shor’s factoring algorithm on the IBM QX5 superconducting processor, successfully factorising the integers N=15, 21, and 35. In contrast, the largest number factorized through a large-scale simulation of Shor’s algorithm was reported in~\cite{Willsch_2023}, where a 39-bit integer was factorized using 2048 GPUs of the JUWELS Booster supercomputer. Although there exist reports claiming the factorization of larger integers on digital quantum computers, recent state-of-the-art reviews such as~\cite{Scholten_2024} and~\cite{Willsch_2025} point out that these implementations often rely on significant oversimplifications of the original problem~\cite{Smolin_2013}.

Another approach to factorisation is to convert the problem into a binary optimisation problem. In this approach, both the number and its factors are treated using their binary representation, to find the combinations that fulfill $N=p \cdot q$. This is a problem suitable for quantum computers. However, in light of their current scalability limitations, \textit{Variational Quantum Algorithms} (VQAs)~\cite{Bharti_2022,Cerezo_2021} have emerged as a promising alternative for exploiting digital-based NISQ devices for solving it, together with other quantum alternatives such as Quantum Annealing. Nevertheless, VQA practical performance can be hindered by several challenges, including hardware noise~\cite{Stilck_2021,Zhang_2022,Bornens_2023}, the appearance of \textit{barren plateaus}~\cite{Larocca_2025}, and the presence of local minima during the optimisation process~\cite{Anschuetz_2022}. More fundamentally, the number of qubits required by most variational formulations typically scales linearly with the number of classical variables involved in the problem~\cite{Weidenfeller_2022}, severely limiting their applicability to large-scale instances.

To address this latter difficulty, \textit{Pauli Correlation Encoding} (PCE)~\cite{Sciorilli_2025} has recently been proposed as a novel encoding framework capable of drastically reducing qubit requirements in binary optimisation problems. By exploiting correlations represented through Pauli operator strings, PCE enables the encoding of $m$ binary variables using only a polynomial number $n$ of qubits, with $n \ll m$. In the original work, PCE achieved competitive results on large-scale \textit{Max-Cut} instances and has since been successfully applied to portfolio optimisation problems~\cite{soloviev2025,Padin_2026}. Moreover, recent benchmarking studies have demonstrated its scalability and effectiveness across several classical combinatorial optimisation problems~\cite{Alonso_2026}.

Motivated by this capability, a fundamental question arises: Can PCE-based algorithms be used for solving the factorisation problem via optimisation techniques with a limited number of resources (quantum or classical)? The present work tries to answer this question and investigates the applicability of a PCE-based algorithm to the factorisation problem. In particular, we explore two alternative strategies for encoding and mapping the problem, thereby reformulating it as a variational optimisation task. The proposed formulations are analysed through numerical experiments, with special attention given to their behaviour under different classical optimizers and to their scalability as the size of the integer to be factorised increases. Beyond evaluating their practical performance, the objective of this study is to assess the feasibility of employing PCE in the context of integer factorisation and to identify the main challenges and limitations associated with such an approach. Rather than claiming any computational advantage over existing factorisation algorithms, this work aims to establish an alternative framework upon which future research may build, providing new insights into the application of quantum and quantum-inspired optimisation techniques to problems in computational number theory.

The manuscript is organised as follows. Section~\ref{sec_2} is divided into two subsections and introduces the factorisation problem by formulating it as an optimisation problem, together with an analysis of both classical and quantum approaches for its solution. Section~\ref{sec_3} describes the methodology used to execute and evaluate the proposed method, specifying the computational components employed, such as the ansatz, the optimiser, the encoding, and the software environment. Section~\ref{sec_4} is dedicated to the implementation of PCE for tackling the factorisation problem and is structured into two subsections presenting different approaches. Section~\ref{sec_4_1} introduces the \textit{Basic approach}, establishing its theoretical formulation and presenting the results obtained under different configurations. Subsequently, Section~\ref{sec_4_2} analyses the \textit{DoTS approach} following the same scheme, highlighting the improvements achieved and providing a more extensive analysis of its performance. Section~\ref{sec_5} provides an overview of the potential of using the PCE, discussing its principal advantages and limitations, while Section~\ref{sec_6} outlines several directions for future research.

\section{Factorisation as optimisation}\label{sec_2}

In this section, we review the formulation of integer factorisation as an optimisation problem. We begin by introducing its classical mathematical formulation and discussing the main optimisation-based approaches that have been proposed in the literature. Subsequently, we examine several variational quantum algorithms designed to address the factorisation problem, highlighting their underlying principles, advantages, and limitations.

\subsection{Classical optimisation methods}\label{sec2_1}

The factorisation problem can be reformulated as an optimisation problem in which the objective is to identify the binary representation of the prime factors that minimises a suitably defined cost function. This perspective enables the application of both continuous and discrete optimisation techniques to search for the factors of a semiprime integer $N$, as originally proposed by \cite{Burges_2002}.

Let the binary representations of the integers be given by
\begin{align*}
N = \sum_{i=0}^{l_N-1} N_i 2^i, \quad
p = \sum_{i=0}^{l_p-1} p_i 2^i, \quad
q = \sum_{i=0}^{l_q-1} q_i 2^i, \quad
\label{eq_org_p_q}
\end{align*}
where $N_i, p_i, q_i \in \{0,1\}$ denote binary variables associated with each bit of $N$, $p$, and $q$, respectively, while $l_N$, $l_p$, and $l_q$ denote the number of bits required to represent each integer. Without loss of generality, assume that $p \geq q$, in which case the following bounds hold:
\begin{equation}
l_p \leq l_N
\quad \text{and} \quad
l_q \leq \left\lceil \frac{l_N}{2} \right\rceil.
\label{eq:bits_estimation}
\end{equation}
The cost function is constructed by expressing the unknown factors in terms of binary variables and translating the multiplication constraint $N = p \cdot q$ into a set of algebraic relations that these variables must satisfy.  To explicitly characterise this multiplication process, each binary column can be analysed independently by considering the partial products contributing to a given bit position, together with the carry terms generated during the standard binary addition procedure. These carry terms correspond to the usual binary remainders propagated to higher-order columns and are represented by auxiliary binary variables $z_{i,j}$ that denote the carry bit propagated from bit position $i$ to bit position $j$.

Table \ref{tab:bit_multiplication} summarises this column-wise decomposition, showing the partial products generated by the pairwise interactions between bits of $p$ and $q$, their associated positional weights, and the auxiliary carry variables involved in the construction.

\begin{table}[h]
\centering
\begin{tabular}{|c|c|c|c|}
\hline
Column $i$ & Partial products & Weight & Auxiliary variables \\
\hline
$0$ & $p_0 q_0$ & $2^0$ & -- \\
$1$ & $p_0 q_1 + p_1 q_0$ & $2^1$ &  carries $z_{j,1}$ \\
$2$ & $p_0 q_2 + p_1 q_1 + p_2 q_0$ & $2^2$ &  carries $z_{j,2}$ \\
$\vdots$ & $\vdots$ & $\vdots$ & $\vdots$ \\
$i$ & $\sum_{j=0}^{i} q_j p_{i-j}$ & $2^i$ &  carries $z_{j,i}$ \\
\hline
\end{tabular}
\caption{Column-wise decomposition of the binary multiplication constraint $p \cdot q = N$.}
\label{tab:bit_multiplication}
\end{table}
This column-wise decomposition can be generalised into a system of polynomial constraints describing the full multiplication process of the form
\begin{equation*}
0 = N_i - \sum_{j=0}^{i} q_j p_{i-j} -\sum_{j=0}^{i} z_{ji} + \sum_{j=1}^{l_p+l_q-1} 2^j z_{i,i+j}, \quad \forall  i \in\{0,\ldots,l_N\},
\end{equation*}
where $z_{i,j} \in \{0,1\}$ denotes the carry bit propagated from bit position $i$ to bit position $j$. To reformulate the previous set of constraints as a binary optimisation problem, a quantity referred to as a clause is introduced for each binary column and defined as
\begin{equation*}
C_i = N_i - \sum_{j=0}^{i} q_j p_{i-j} - \sum_{j=0}^{i} z_{ji} + \sum_{j=1}^{l_p+l_q-1} 2^j z_{i,i+j},
\quad \forall , i \in \{0,\ldots,l_N\}.
\end{equation*}
The factorisation problem can then be recast as the minimisation of the global cost function
\begin{equation}
\mathcal{L} = \sum_{i=0}^{l_N} C_i^2,
\label{eq_L_clauses}
\end{equation}
whose global minimum corresponds to the simultaneous satisfaction of all binary multiplication constraints and therefore yields a valid factorisation of the integer $N$.

Several preprocessing techniques have been proposed to reduce the number of variables involved in the optimisation formulation by classically simplifying a subset of the binary constraints~\cite{Xu_2012,Dattani_2014}. In particular, some approaches iteratively analyse the clauses and directly solve for variables that can be efficiently determined beforehand, thereby reducing the overall size of the optimisation problem. 

However, although these preprocessing strategies can significantly reduce the dimensionality of the search space, they do not eliminate the intrinsic computational hardness of the factorisation problem. From an algebraic perspective, as discussed in \cite{Burges_2002}, the multiplication constraints in \eqref{eq_L_clauses} are highly coupled, since each term contains multiple crossed interactions between binary variables associated with both factors. This strong variable correlation significantly limits the effectiveness of sequential solving strategies.

At the same time, the computational difficulty of this formulation can also be understood by analysing the structure of the corresponding optimisation landscape. In particular, the main source of hardness arises from its combinatorial nature, as the number of local minima grows rapidly with the dimension of the search space, according to the result in \textit{Theorem} 14 of~\cite{Burges_2002}, based on~\cite{Hardy_1917}. This number of local minima makes it increasingly difficult for local optimisation methods to identify the global minimum.

Alternative formulations of the factorisation problem have also explored optimisation-based perspectives beyond the direct binary encoding described before. In particular, Schnorr~\cite{Schnorr_1997, Vera_2010} introduced a lattice-based formulation in which integer factorisation is reduced to a geometric search problem over high-dimensional lattices. The central idea consists of constructing a lattice whose vectors encode algebraic relations associated with the target integer. Factorisation can then be approached by solving lattice problems such as the \textit{Shortest Vector Problem} (SVP) or the \textit{Closest Vector Problem} (CVP)~\cite{Schnorr_2013}, where the objective is to identify vectors satisfying specific structural properties that reveal non-trivial divisors of the composite integer. 

Taken together, these developments highlight that integer factorisation can be reformulated through fundamentally different mathematical frameworks, where optimisation and structured search strategies provide alternative computational approaches beyond traditional number-theoretic algorithms.

\subsection{Quantum optimisation methods}\label{sec2_2}

The prospect of efficiently solving the factorisation problem has made this task one of the central applications of quantum computing since the introduction of Shor’s algorithm. However, the impracticality of implementing fault-tolerant quantum algorithms on current NISQ hardware has motivated increasing interest in alternative quantum variational approaches.

The most widely studied framework is \textit{Variational Quantum Factoring} (VQF)~\cite{Anschuetz_2018}, which reformulates integer factorization as an Ising-type optimization problem, following the perspective introduced in~\cite{Burges_2002}. In the original proposal, the authors introduce a classical preprocessing stage where the factorization equations are simplified over Boolean variables in order to reduce the number of qubits required for the Hamiltonian encoding. The resulting Hamiltonian is then solved using QAOA, seeking an approximate ground state from which the prime factors can be recovered.

In~\cite{Karamlou_2021}, this framework was implemented on a superconducting quantum processor, analyzing the trade-off between qubit count and circuit depth in order to identify circuit configurations that maximize the success probability of biprime factorization. The reported results demonstrate the factorization of 13, 17, and 40-bit integers using 3, 4, and 5 qubits, respectively, while also examining the impact of different noise sources on the algorithm’s performance.

More recently, alternative optimization strategies based on the \textit{Variational Quantum Eigensolver} (VQE)~\cite{Peruzzo_2014} have also been explored within the VQF framework. In~\cite{Sobhani_2025}, the authors evaluate a VQE-based approach without relying on preprocessed arithmetic simplifications, demonstrating the factorization of 8 and 20-bit integers using 9 and 27 qubits, respectively, on both IBM quantum hardware and classical simulation. Later,~\cite{Phan_2022} compared QAOA and VQE as optimization strategies for VQF, reporting improved performance when employing VQE for ground state preparation.

Beyond direct variational formulations of the factorization problem such as the previous ones, recent research has also explored quantum and quantum-inspired approaches that operate within classical factorization frameworks. 

In~\cite{Yan_2022}, a 48-bit integer was factorized using 10 superconducting qubits through a universal quantum algorithm derived from Schnorr’s factorization method. This approach requires only sublinear quantum resources, with the number of qubits scaling as $\mathcal{O}(m/\log m)$ for an $m$-bit integer $N$. In particular, the authors employ QAOA to optimize the most computationally demanding component of the algorithm, improving the overall efficiency of the factorization process.

In~\cite{Hong_2025}, an 80-bit integer was factorized using D-Wave’s hybrid quantum-classical architecture. Building upon the same Schnorr-based factorization framework, the authors leverage quantum annealing techniques to accelerate critical computational steps, enabling a more efficient search process and improving the overall performance of RSA factorization.

More recently,~\cite{Tesoro_2026} reports the factorization of a 100-bit integer using Tensor Network Schnorr’s Sieving (TNSS), a quantum-inspired approach that reformulates Schnorr’s sieving procedure using tensor network methods. The authors provide numerical evidence suggesting polynomial scaling of computational resources with the bit-length of the biprime.

\section{Methodology}\label{sec_3}

This section describes the main technical aspects that characterise the experimental framework and specifies the conditions under which the simulations and experiments considered have been conducted.

\begin{figure}[H]
\centering
\includegraphics[width=0.75\textwidth]{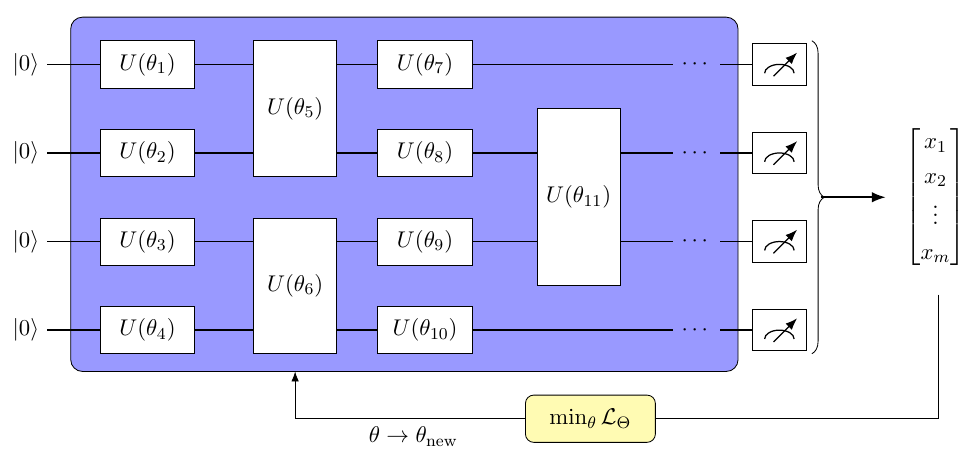}
\caption{Schematic of the quantum circuit used to construct the PQC.}
\label{fig:esquemansatz}
\end{figure}

Following the study developed in~\cite{Alonso_2026}, based on the original PCE proposal, the parameterised quantum circuits employed in this work are constructed using the same problem-agnostic, hardware-efficient brickwork ansatz shown in Figure~\ref{fig:esquemansatz}. 

The ansatz consists of alternating layers of parameterised single- and two-qubit rotation gates. Single-qubit layers are formed by \texttt{RX}, \texttt{RY}, and \texttt{RZ} rotations applied cyclically, whereas entangling layers are built from nearest-neighbour \texttt{RXX}, \texttt{RYY}, and \texttt{RZZ} gates arranged in an alternating brickwork pattern. Each gate introduces a single variational parameter, allowing the expressive power of the circuit to be increased in a controlled manner as its depth grows~\cite{Schuld_2021}. Additionally, the circuit depth scales as $\mathcal{O}\left(m^{1-\frac{1}{k}}\right)$, where $m$ denotes the number of variables and $k$ the compression order.

Following the methodology adopted in~\cite{Alonso_2026}, \textit{Differential Evolution} (DE) is employed as the baseline optimiser throughout this work. This gradient-free evolutionary strategy was selected in the original study due to its reported robustness against the presence of local minima~\cite{Failde2023}.

Besides DE, we consider two additional optimisers. The first of these is \textit{Particle Swarm Optimisation} (PSO)~\cite{Sengupta_2018,Jones_2024}, a bio-inspired, population-based, gradient-free optimisation algorithm. It operates by evolving a set of candidate solutions, referred to as particles, each characterised by position and velocity. In our case, position refers to circuit parameters, and, during the optimisation process, particle positions are updated according to their respective velocities. While no guarantee of global optimality can be provided, PSO has demonstrated strong empirical performance across a wide variety of optimisation problems. In this work, the PSO implementation was adapted from the open-source repository in~\cite{sotostzam_pso}.


The other optimiser considered is \textit{Quantum-Behaved Particle Swarm Optimisation} (QPSO)~\cite{Yang_2004}, a closely related variant that replaces the velocity-based dynamics of classical PSO with a quantum-inspired search mechanism. In this framework, particle positions are sampled according to a probabilistic model governed by a quantum dispersion parameter, allowing a broader exploration of the search space while preserving the collective learning behaviour characteristic of swarm-based methods. In this work, the implementation provided in~\cite{ngroup_qpso} is employed. Since this implementation is based on the delta potential well formulation originally proposed in~\cite{Sun_2004}, it will be referred to throughout this work as \textit{Quantum Delta Particle Swarm Optimisation} (QDPSO).

The specific parameters used for the optimizer DE are indicated in Table~\ref{tab:DEparameters}.

\begin{table}[H]
\centering
\begin{tabular}{|c|c|}
\hline
\textbf{Parameter} & \textbf{Value} \\
\hline
strategy & best1exp \\
maxiter & 10000 \\
popsize & 10 \\
tol & 0 \\
mutation & $(0.5,\,1)$ \\
recombination & 0.7 \\
polish & False \\
\hline
\end{tabular}
\caption{DE configuration parameters.}\label{tab:DEparameters}
\end{table}

The specific parameters used for the optimizer PSO are indicated in Table~\ref{tab:PSOparameters}.

\begin{table}[H]
\centering
\begin{tabular}{|c|c|}
\hline
\textbf{Parameter} & \textbf{Value} \\
\hline
num particles & 100 \\
maxiter & 10000 \\
\hline
\end{tabular}
\caption{\textit{PSO} configuration parameters.}\label{tab:PSOparameters}
\end{table}

The specific parameters used for the optimizer QDPSO are indicated in Table~\ref{tab:QDPSOparameters}.

\begin{table}[H]
\centering
\begin{tabular}{|c|c|}
\hline
\textbf{Parameter} & \textbf{Value} \\
\hline
num particles & 100 \\
maxiter & 10000 \\
Quantum dispersion parameter ($g$) & 0.96 \\
\hline
\end{tabular}
\caption{\textit{QDPSO} configuration parameters.}\label{tab:QDPSOparameters}
\end{table}

Following the observations reported in~\cite{Alonso_2026} and considering that all optimisation procedures employed in this work are gradient-free, the sign function was adopted instead of the hyperbolic tangent relaxation originally proposed in PCE. While the latter provides a smoother approximation of the binary variables, the reconstructed candidate factors would be real and $N = p \cdot q$ has infinite solutions for $p, q \in \R$, hindering the optimisation process, as the only valid solution corresponds to integer values of $p$ and $q$. Furthermore, the use of the hyperbolic tangent relaxation requires the introduction and tuning of the hyperparameter $\alpha$ to identify suitable operating regimes. The optimal values of these parameters may vary significantly across problem instances and parameter initialisations, making their calibration a non-trivial and computationally demanding task. For similar reasons, the regularisation term controlled by $\beta$ was not included in the formulations considered in this work. Besides introducing an additional hyperparameter, previous results indicate that its contribution to solution quality is often limited, with comparable performance frequently obtained for $\beta = 0$.


The software used is indicated in Table~\ref{tab:software_versions}.

\begin{table}[H]
\centering
\begin{tabular}{|c|c|}
\hline
\textbf{Software} & \textbf{Version} \\
\hline
Python & 3.11.9 \\
NumPy & 1.26.4 \\
SciPy & 1.13.0 \\
Pandas & 2.2.2 \\
NetworkX & 3.3 \\
Qiskit & 1.2.4 \\
Qiskit Aer & 0.15.1 \\
\hline
\end{tabular}
\caption{Software environment and versions used in the experiments.}
\label{tab:software_versions}
\end{table}

It should be noted that, due to the stochastic nature of the aforementioned optimisers, the circuit parameters were initialised randomly within the interval $[0, 2\pi)$. Additionally, exact simulations were carried out using \textit{QiskitAer}, with the \textit{AerSimulator} emulator; in both cases, the \textit{Statevector} and \textit{Matrix Product State} default methods were employed.

\section{Factorisation through PCE}\label{sec_4}

In this section, we present two different strategies for encoding the factorisation problem using PCE. Both approaches reformulate factorisation as a binary optimisation problem, although they differ in the way the factorisation constraints are represented within the cost function. Their main characteristics are analysed and compared, highlighting the advantages and drawbacks of each formulation in terms of implementation, scalability, and optimisation behaviour.

Unlike some of the optimisation-based formulations of integer factorisation introduced in the previous sections, which explicitly model the binary multiplication process through carry variables and instance-dependent arithmetic constraints, the approaches considered here seek to preserve a more generic setting. The cost functions are constructed directly from the binary representations of the factors and the target integer, avoiding the introduction of auxiliary variables and reducing the dependence of the formulation on the particular integer being factored.

\subsection{Basic approach}\label{sec_4_1}

The first formulation considered, referred to as the \textit{Basic approach}, provides a direct encoding of the factorisation problem. The optimisation variables are associated with the binary representations of the unknown factors $p$ and $q$, so that finding a solution to the optimisation problem corresponds to determining the bit strings whose product equals the target integer $N$.

To construct this formulation, it is first necessary to determine the number of bits required to represent the integer $N$. This quantity will be given by
\begin{equation*}
l_N = \lfloor \log_2(N) \rfloor + 1.
\end{equation*}
Once this value is known, upper bounds for the bit lengths of the factors, denoted by $l_p$ and $l_q$, can be established. Following the analysis in~\ref{sec_4_2_1}, and assuming without loss of generality that $p \geq q$, we can use the bounds in \eqref{eq:bits_estimation}.
Although in the most general setting, no information about the factors is assumed a priori, the least significant bit of both $p$ and $q$ can be fixed to 1. Indeed, since prime factors greater than 2 are necessarily odd, their binary representations must end in a 1. Hence, we define
\begin{equation*}
\vec{x}_p = (1,x_1,\dots,x_{l_p-1})
\quad \text{and} \quad
\vec{x}_q = (1,x_{l_p},\dots,x_{l_p+l_q-2}),
\label{eq_binary_p_q}
\end{equation*}
as the binary decision variables of the problem. This formulation requires the encoding of
\begin{equation*}
m = (l_p-1) + (l_q-1)
= \Bigg\lceil \frac{3l_N}{2} \Bigg\rceil - 2
\end{equation*}
variables. Therefore, $m$ grows linearly with the bit-length of the integer to be factorised, or equivalently, logarithmically with $N$.

Each assignment of $\vec{x}_p$ and $\vec{x}_q$ uniquely defines two candidate integers, denoted by $x_p$ and $x_q$, obtained by interpreting the corresponding bit strings in base 2,
\begin{equation*}
x_p = 1 + \sum_{i=1}^{l_p-1} x_i2^i 
\quad \text{and} \quad
x_q = 1 + \sum_{j=1}^{l_q-1}x_{l_p+j-1}~ 2^j .
\end{equation*}
The goal is then to construct a cost function whose minimum is attained when these candidate integers coincide with the prime factors of $N$, that is,
\begin{equation*}
x_p = p
\quad \text{and} \quad
x_q = q.
\end{equation*}
To achieve this, a measure quantifying the discrepancy between the target integer $N$ and the candidate product $x_p\cdot x_q$ associated with a given assignment is first required. Since both quantities admit a base-2 representation, we employ the Hamming distance as a natural metric to quantify the discrepancy between their corresponding bit strings. More precisely, given two integers $u$ and $v$, let
\begin{equation*}
(u_0,u_1,\dots,u_{s-1})
\quad \text{and} \quad
(v_0,v_1,\dots,v_{s-1}),
\end{equation*}
denote their corresponding binary representations using a common length $s$. Their Hamming distance is then defined as
\begin{equation*}
d_H(u,v)=\sum_{i=0}^{s-1}|u_i-v_i|.
\end{equation*}
Additionally, we introduce an indicator function that takes the value $1$ whenever two integers $u$ and $v$ are equal, and $0$ otherwise:
\begin{equation*}
\mathbb{I}(u,v)=
\begin{cases}
1 \quad \text{if} \quad u=v\\
0 \quad \text{if} \quad u\neq v.
\end{cases}
\end{equation*}
The cost function is then defined as
\begin{equation*}
\mathcal{L}=
\Big(d_H(N,x_p\cdot x_q)\Big)^2
+ \lambda \cdot \Big(
\mathbb{I}(N, x_p)
+ \mathbb{I}(1,  x_p)
+ \mathbb{I}(1,  x_q)
\Big),
\label{eq: loss_fun_org}
\end{equation*}
The first term computes the squared Hamming distance between the binary representations of the target integer $N$ and the candidate product $x_p \cdot 
x_q$, thereby penalising the number of bit positions in which both representations differ. Consequently, minimising this term drives the optimisation process towards assignments whose associated candidate integers satisfy the factorisation condition of $N$.

The second term is introduced to prevent trivial solutions. In particular, without additional constraints, the optimisation may converge to the trivial factorisation $N = N \cdot 1$, which formally satisfies the factorisation condition but is of no practical interest. To discourage such outcomes, a penalty term weighted by the coefficient $\lambda$ is incorporated through the indicator function. In general, $\lambda$ is chosen to be equal to the integer being factorised, such that its order of magnitude provides an appropriate upper bound for the penalty contribution.

\subsubsection{Experiments and results}\label{sec_4_1_1}

In this section, we analyse the performance of the proposed approach by comparing different compression orders $k$ of the PCE-based algorithm as the bit-length of the integers to be factored increases. In particular, the semiprime instances considered in this study were generated randomly using the \textit{RSA Challenge Generator}~\cite{bigprimes_rsa}, ensuring a consistent and unbiased test across all experiments.

\begin{figure}[H]
    \centering
     \subfloat[$k\in\{1,2,3,4\}$]{\includegraphics[width = 0.51\textwidth]{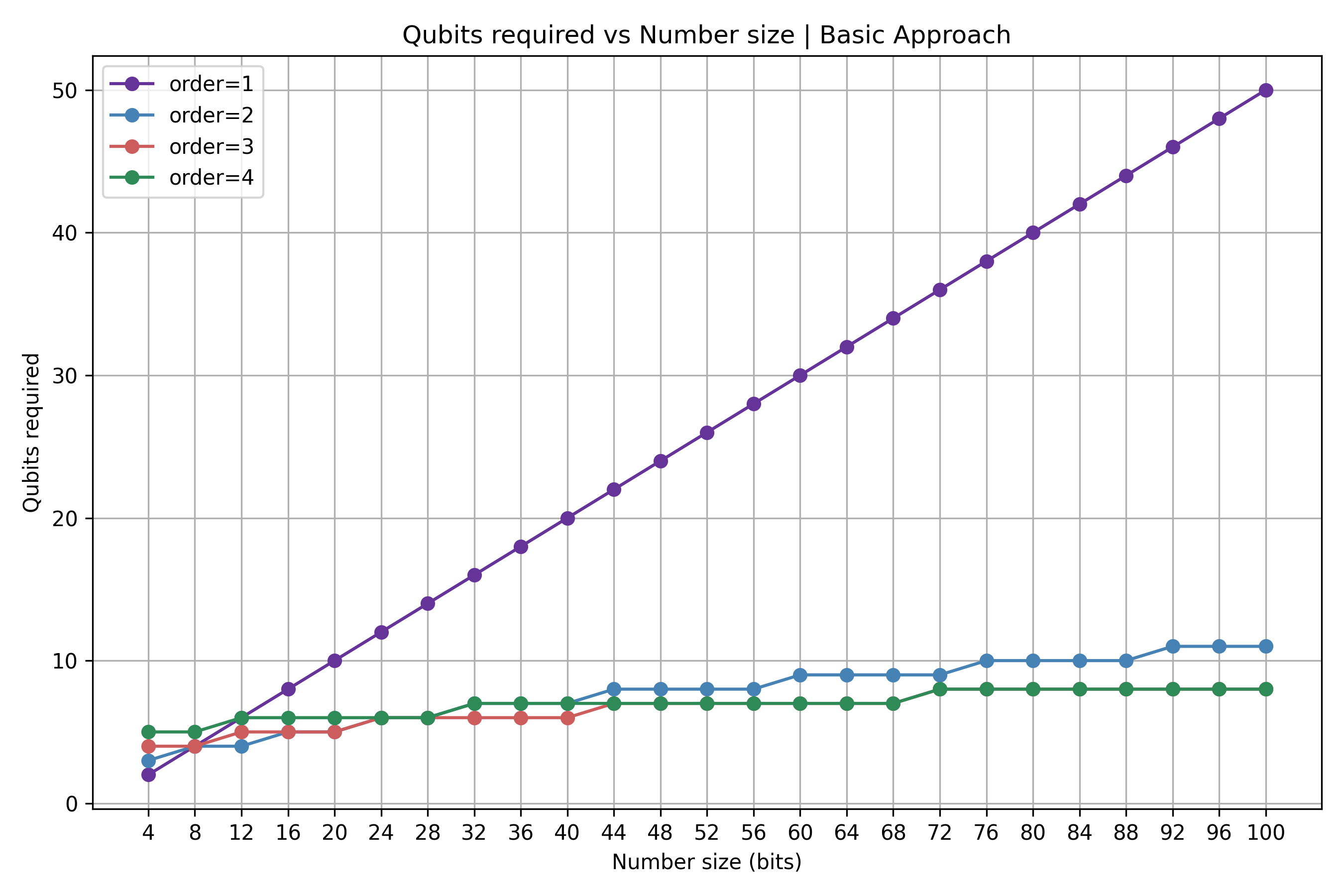}}
     \subfloat[$k\in\{2,3,4\}$]{\includegraphics[width = 0.51\textwidth]{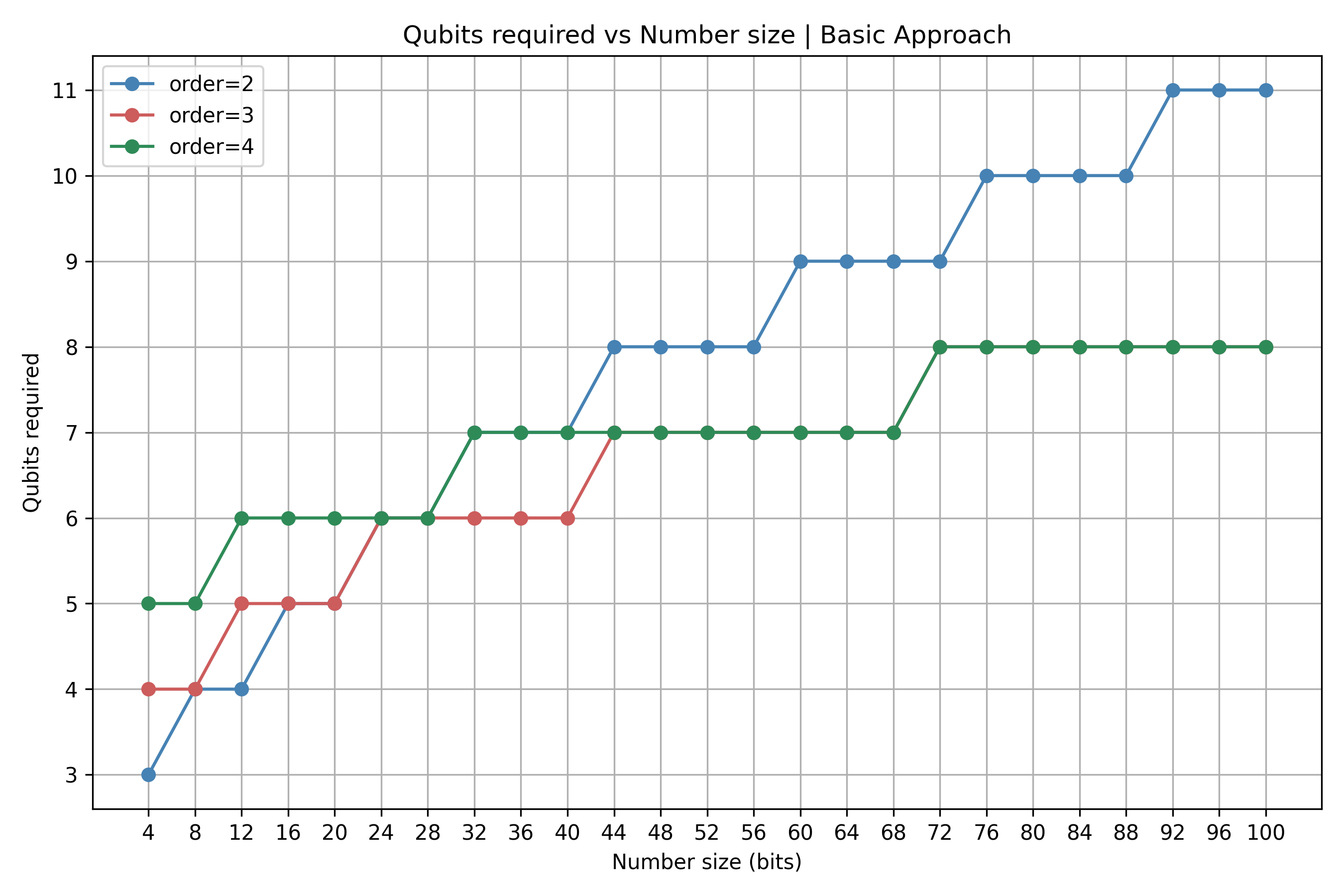}}
    \caption{Scaling of the number of qubits as a function of the number of bits.}
    \label{fig: Fact_org_num_qubits}
\end{figure}

\begin{figure}[H]
    \centering
     \subfloat[Depth for $k\in\{2,3,4\}$]{\includegraphics[width = 0.51\textwidth]{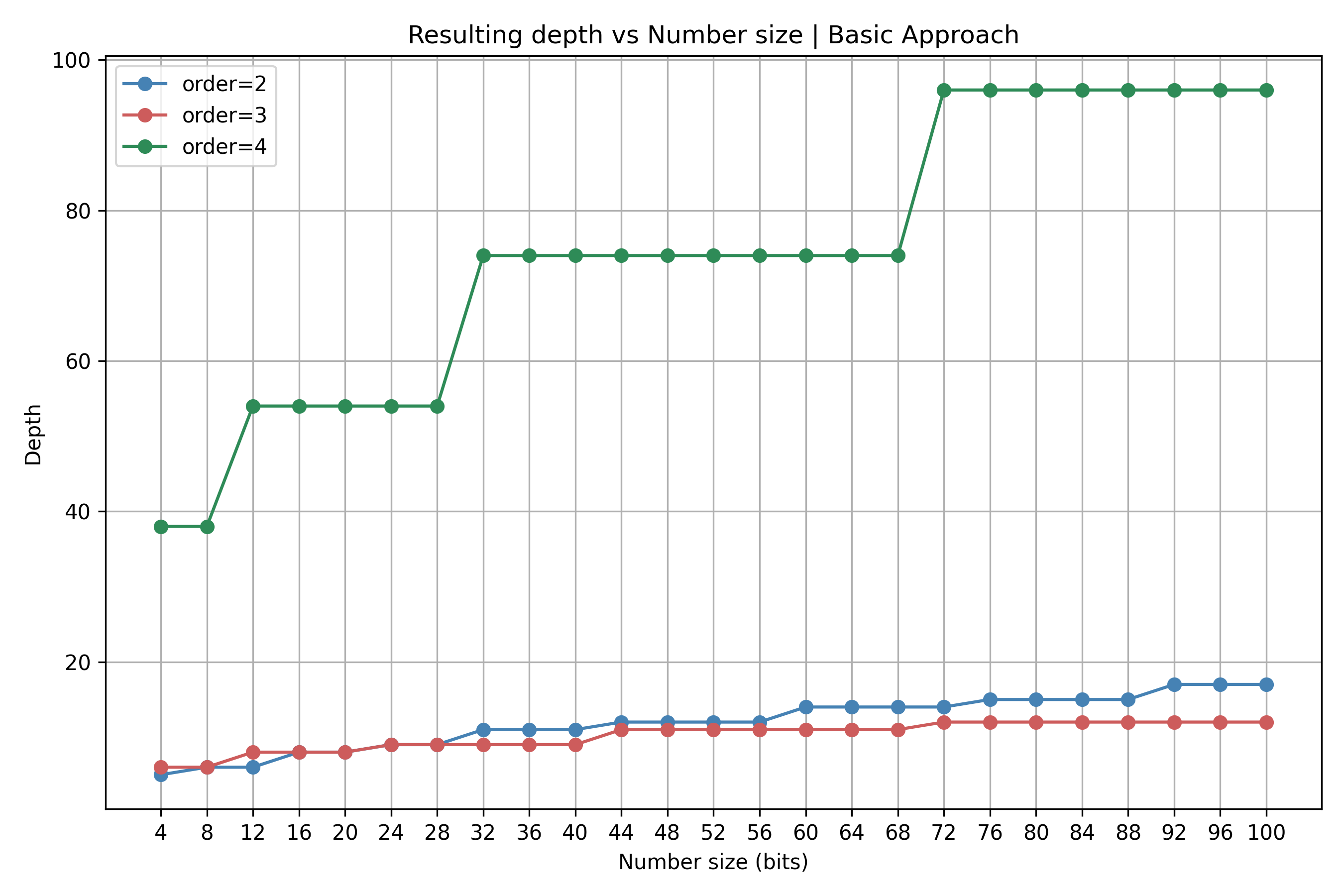}}
     \subfloat[Number of params for $k\in\{2,3,4\}$]{\includegraphics[width = 0.51\textwidth]{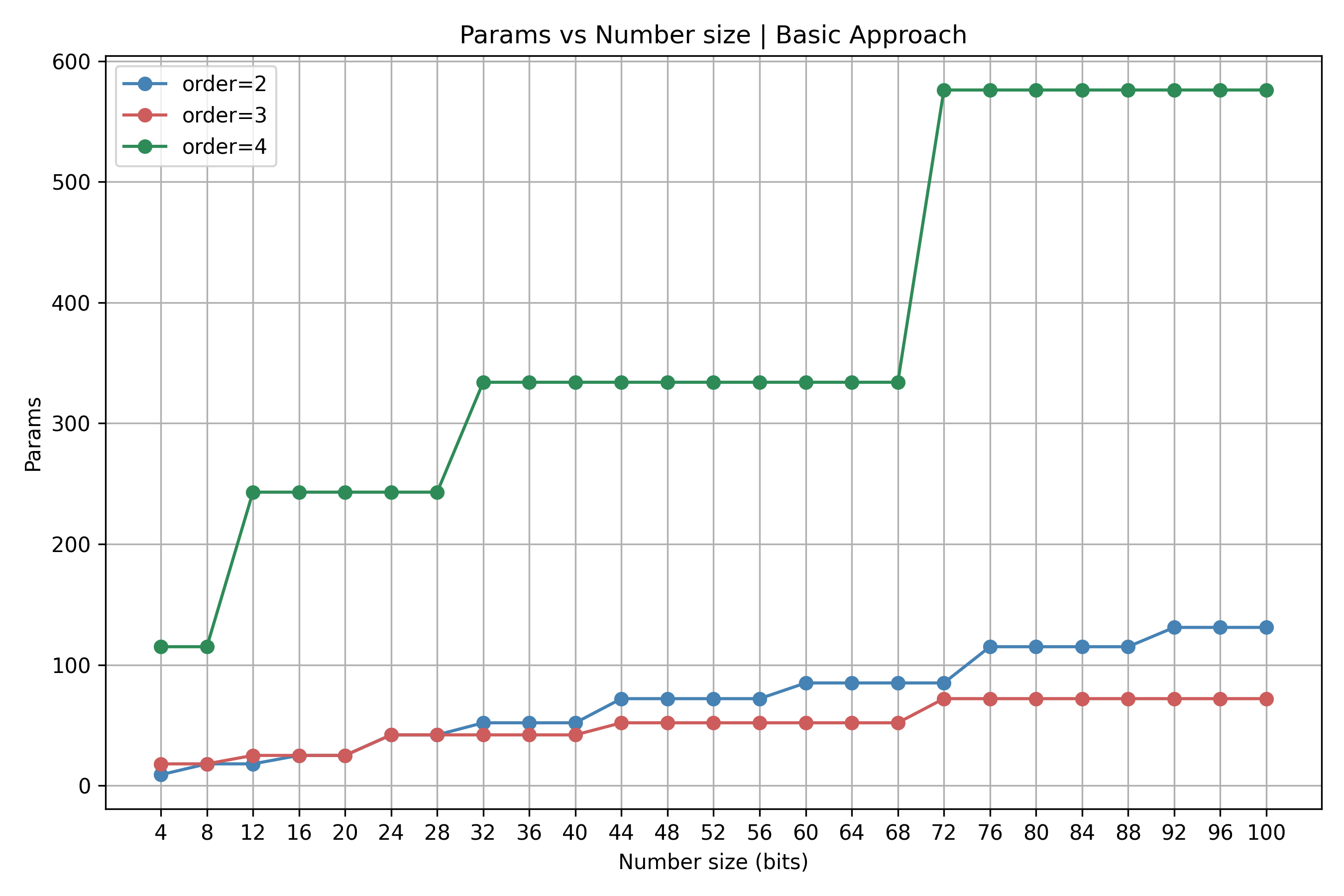}}
    \caption{Scaling of the ideal circuit depth and resultant number of parameters as a function of the number of bits.}
    \label{fig: Fact_org_resources}
\end{figure}

Firstly, the number of qubits, circuit depth (measured as the number of ansatz layers), and variational parameters resulting from the PCE configuration employed in this approach are shown in \textit{Figures}~\ref{fig: Fact_org_num_qubits} and~\ref{fig: Fact_org_resources}, respectively.

Under the chosen configuration, the compression order $k$ plays a crucial role in determining the number of variational parameters of the resulting circuit. Increasing $k$ generally leads to a larger parameter space and, consequently, to a more expressive optimisation model, potentially improving its ability to identify high-quality solutions provided that the associated resource requirements remain manageable.

This behaviour is also reflected in the factorisation experiments shown below. Figure~\ref{fig:Fact_org_results} reports both the factorisation success rate and the proportion of trivial solutions as functions of the bit-length of the semiprime instances, for different compression orders $k$ and optimisers. In particular, semiprime instances ranging from 17 to 25 bits were considered, and a clear dependence can be observed on both the compression order and the size of the integer. As illustrated in the figure, the success rate generally increases with $k$, while decreasing as the bit-length of the integer grows. Among the tested configurations, $k=4$ consistently achieves the best overall performance, outperforming both $k=2$ and $k=3$ on the smaller instances and remaining the only configuration capable of successfully factoring some of the larger numbers, even though with a considerably reduced success rate.

The results also reveal a strong dependence on the choice of optimiser. In \textit{Figure}~\ref{fig:Fact_org_results}, DE is the only optimiser that consistently achieves successful factorisations across the considered instances. By contrast, both PSO and QDPSO exhibit substantially lower success rates, indicating that the optimisation landscape associated with this formulation is considerably more challenging for swarm-based optimisation methods.

Despite these results, it can be seen that the success rate decreases significantly for instances larger than approximately 25 bits. This observation suggests that the cost function employed in the basic formulation may not fully capture the underlying correlations among the variables involved in the factorisation problem.

\newpage
~
\vspace{3ex}

\begin{figure}[H]
    \centering
    \subfloat[ $k=2$]{%
        \includegraphics[width=1.0\textwidth]{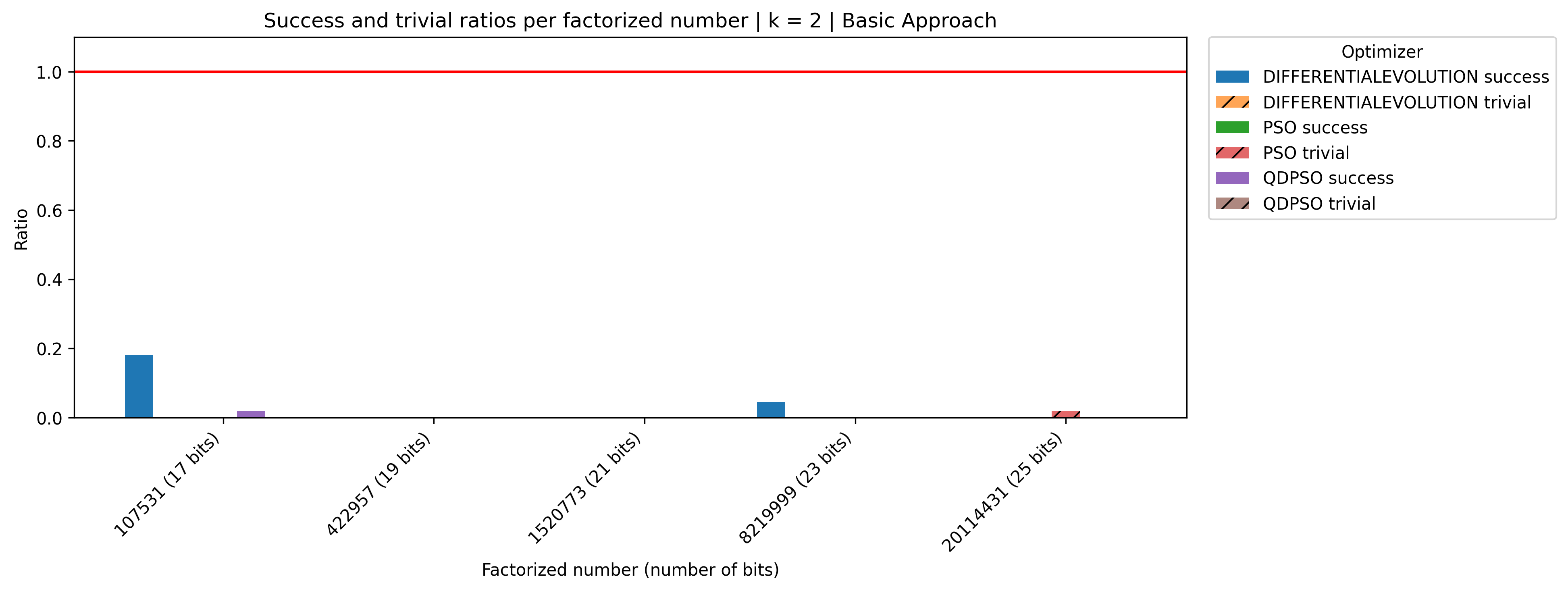}%
    }\\[1ex]
    \subfloat[ $k=3$]{%
        \includegraphics[width=1.0\textwidth]{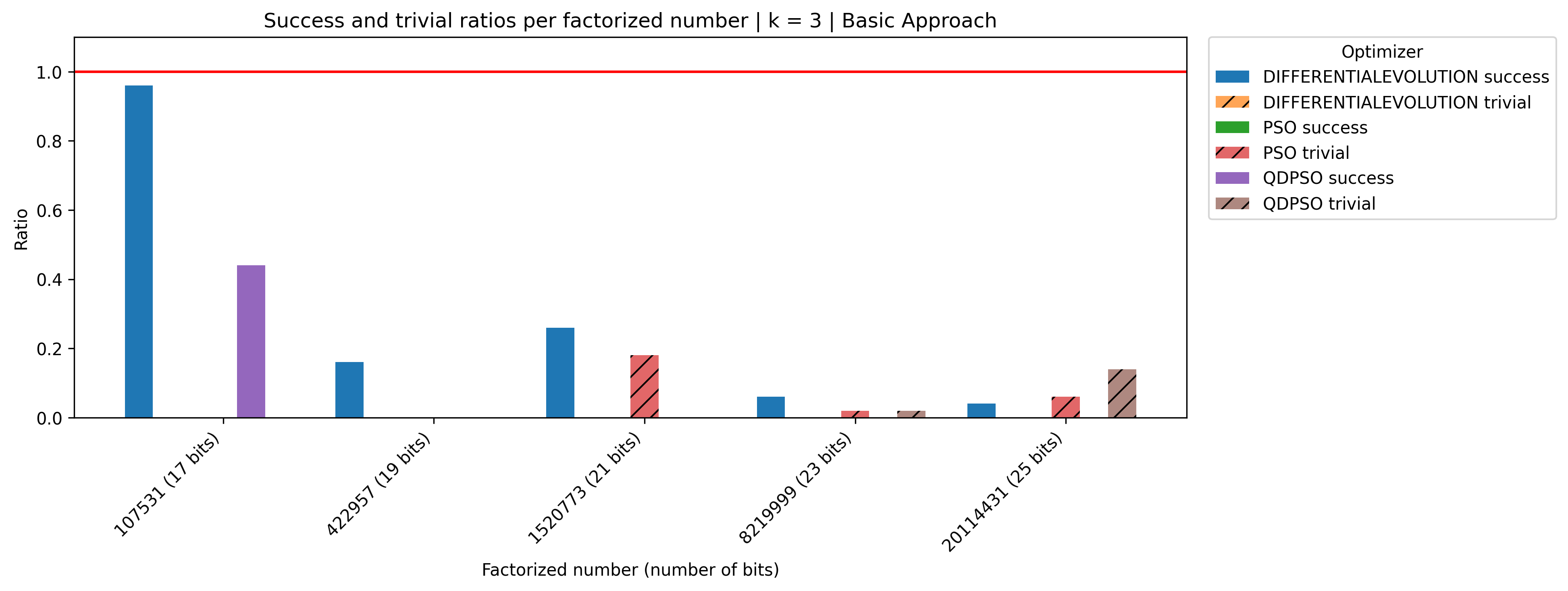}%
    }\\[1ex]
    \subfloat[ $k=4$]{%
        \includegraphics[width=1.0\textwidth]{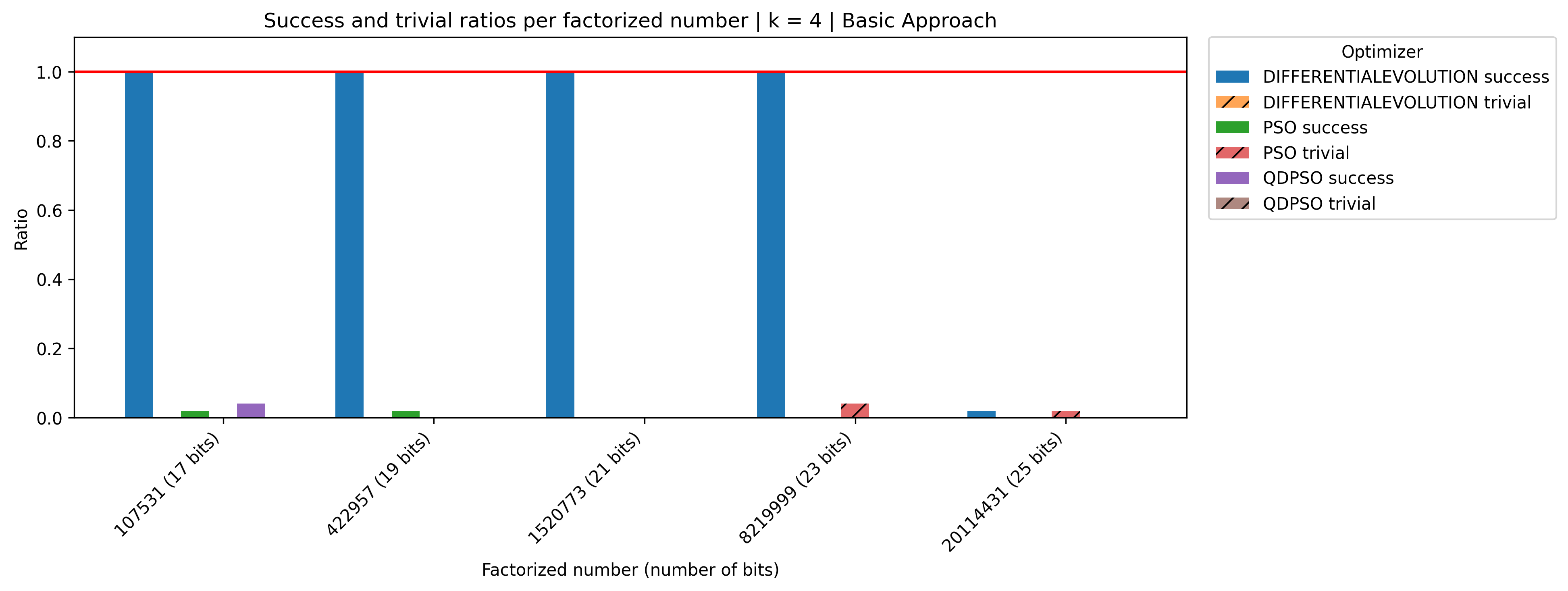}%
    }
    \caption{Factorisation success and trivial solution rates as a function of the number of bits, for different compression orders $k$ and optimisers, over 50 random initialisations.}
    \label{fig:Fact_org_results}
\end{figure}

\subsection{DoTS approach}\label{sec_4_2}

Motivated by the limitations observed in the first approach, an alternative mapping of the factorisation problem into the PCE framework was investigated with the aim of defining a more complete cost function and, consequently, improving the probability of recovering the correct factorisation. This second formulation is based on a classical arithmetic property that appears in several factorisation algorithms and whose most elementary form is already present in Fermat's method~\cite{Gupta_2009}. Rather than searching directly for the factors $p$ and $q$, the problem is reformulated as the search for two integers $a$ and $b$ satisfying a difference-of-squares relation associated with the target integer $N$. For this reason, the proposed formulation will be referred to as the \textit{Difference of Two Squares approach} (DoTS).

Assume that there exist integers $a,b\in\mathbb{Z}$ satisfying
\begin{equation*}
a^2 \equiv b^2 \pmod N.
\end{equation*}
This congruence can be rewritten as
\begin{equation*}
a^2-b^2 \equiv 0 \pmod N
\quad \Longleftrightarrow \quad
(a-b)(a+b) \equiv 0 \pmod N.
\end{equation*}
Hence, $N$ divides the product $(a-b)(a+b)$. If $N$ is a semiprime, this implies that its prime factors must be distributed between the terms $(a-b)$ and $(a+b)$. Consequently, non-trivial factors of $N$ can be recovered through the greatest common divisors
\begin{equation*}
p=\gcd(a-b,N)
\quad \text{and} \quad 
q=\gcd(a+b,N).
\end{equation*}

However, not every solution of
\begin{equation*}
a^2 \equiv b^2 \pmod N
\end{equation*}
yields a non-trivial factorisation. Indeed, certain solutions lead only to the trivial divisors, making it necessary to identify congruences that produce non-trivial greatest common divisors.

In particular, the congruences
\begin{equation*}
a \equiv b \pmod N
\quad \text{or} \quad
a \equiv -b \pmod N
\end{equation*}
lead only to trivial factorisations. Indeed:
\begin{itemize}
\item If $a \equiv b \pmod N$, then $N \mid (a-b)$ and therefore $\gcd(a-b,N)=N$.

\item If $a \equiv -b \pmod N$, then $N \mid (a+b)$ and therefore $ \gcd(a+b,N)=N$.
\end{itemize}

Consequently, the search must be restricted to non-trivial solutions satisfying
\begin{equation*}
a \not\equiv \pm b \pmod N.
\end{equation*}
In this case, $(a-b)$ can be written as
\begin{equation*}
(a-b)=\gamma N+r
\end{equation*}
for some $\gamma \in \N$ and $1\leq r < N$. By the invariance of the greatest common divisor under the addition of multiples of $N$, it follows that
\begin{equation*}
\gcd(a-b,N)=\gcd(r,N).
\end{equation*}

Therefore, the existence of a non-trivial factorisation can be characterised in terms of the arithmetic relation between $r$ and $N$:
\begin{itemize}
\item If $\gcd(r,N)=1$, then only a trivial divisor is obtained.

\item If $1<\gcd(r,N)<N$, then a non-trivial factor of $N$ is recovered.
\end{itemize}

Hence, rather than searching directly for the factors of $N$, the problem can be reformulated as the search for non-trivial solutions of
\begin{equation*}
a^2 \equiv b^2 \pmod N,
\end{equation*}
for which the corresponding remainder $r$ shares a non-trivial common divisor with $N$. From an optimisation perspective, the objective is therefore to identify suitable pairs $(a,b)$ satisfying this congruence while avoiding the trivial cases discussed above. Once such a pair is found, non-trivial factors of $N$ can be recovered through the corresponding greatest common divisors. In this way, the factorisation is obtained indirectly through the arithmetic structure of the solution rather than by explicitly searching for the factors themselves.

To construct this formulation, we restrict the search to integers satisfying $0 \leq a,b < N$. This assumption does not entail any loss of generality, since the congruence
\begin{equation*}
a^2 \equiv b^2 \pmod N
\end{equation*}
depends only on the residue classes of $a$ and $b$ modulo $N$, and therefore every solution admits an equivalent representative within this interval. From a practical perspective, this restriction is also important to prevent unbounded growth in the number of optimisation variables, which would severely compromise the scalability of the method. Since both $a$ and $b$ are bounded by $N$, the number of bits required to represent them, denoted by $l_a$ and $l_b$, satisfies
\begin{equation*}
l_a,l_b \leq l_N.
\end{equation*}
Henceforth, we define
\begin{equation}
\vec{x}_a=(x_0,x_1,\dots,x_{l_a-1})
\quad\text{and}\quad
\vec{x}_b=(x_{l_a},x_{l_a+1},\dots,x_{l_a+l_b-1})
\label{eq_binary_a_b}
\end{equation}
as the binary decision variables of the problem. This formulation requires the encoding of
\begin{equation*}
m=l_a+l_b=2l_N
\end{equation*}
binary variables, which is larger than in the basic formulation. Nevertheless, $m$ still grows linearly with the bit-length of $N$, or equivalently, logarithmically with $N$.

Each assignment of $\vec{x}_a$ and $\vec{x}_b$ uniquely defines two candidate integers, denoted by $x_a$ and $x_b$, obtained by interpreting the corresponding bit strings in base 2. The candidate factors are then recovered through
\begin{equation*}
x_p=\gcd(x_a-x_b,N)
\quad\text{and}\quad
x_q=\gcd(x_a+x_b,N).
\end{equation*}
The objective is to construct a cost function whose minimum is attained when the induced candidate integers satisfy
\begin{equation*}
x_a = a
\quad \text{and} \quad
x_b = b,
\end{equation*}
thereby ensuring that the previous relations recover a non-trivial factorisation of $N$, i.e, $x_p=p$ and $x_q=q$. Otherwise, assignments that do not correspond to valid solutions yield the trivial factorisation.

The cost function is then defined as the sum of two distinct contributions. The first contribution enforces the arithmetic conditions that the integers $x_a$ and $x_b$ must satisfy to recover a valid non-trivial factorisation:
\begin{align*}
\mathcal{L}_{ab}
=
&\Big(\left[x_a^2 - x_b^2\right]\pmod N\Big)^2  + \\ 
&+\lambda \cdot \Bigg(\mathbb{I}(x_a,x_b) + \mathbb{I}(x_a,-x_b)
+\mathbb{I}\Big(\left[x_a^2 - x_b^2\right],0\Big)
+\mathbb{I}\Big(\left[x_a-x_b\right]\pmod N,0\Big)
\Bigg).
\end{align*}
The second contribution directly evaluates the factorisation induced by the candidate factors $x_p$ and $x_q$ through the factorisation criterion introduced in the previous formulation:
\begin{equation*}
\mathcal{L}_{pq}=
\Big(d_H(N,x_p\cdot x_q)\Big)^2
+ \lambda \cdot \Big(
\mathbb{I}(N, x_p)
+ \mathbb{I}(1,  x_p)
+ \mathbb{I}(1,  x_q)
\Big).
\end{equation*}
The total cost function is then defined as
\begin{equation*}
\mathcal{L} = \mathcal{L}_{ab} + \mathcal{L}_{pq}.
\label{eq:loss_fun_extended}
\end{equation*}
As in the previous formulation, $\lambda$ is typically chosen to be equal to the integer being factorised.

\subsubsection{Experiments and results}\label{sec_4_2_1}


In this section, we build upon the observations made in the previous section by first comparing different compression orders $k$ in order to identify the most effective configuration for the proposed approach. Once this reference configuration has been selected, we analyse its performance in greater detail by extending the study to larger semiprime instances, with the aim of assessing its scalability and determining the range of bit lengths for which successful factorisations can still be achieved.

\begin{figure}[H]
    \centering
     \subfloat[$k\in\{1,2,3,4,5\}$]{\includegraphics[width = 0.51\textwidth]{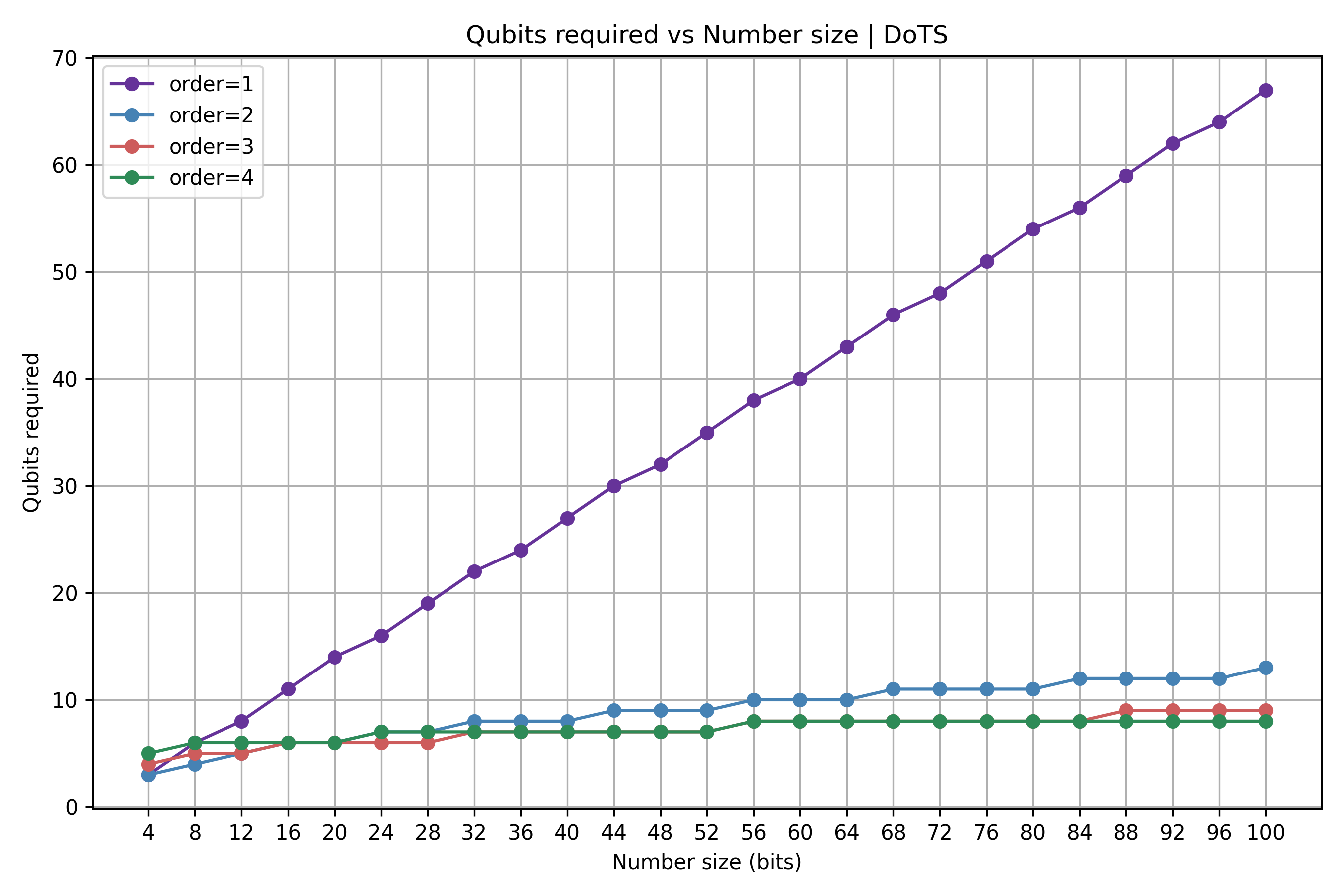}}
     \subfloat[$k\in\{2,3,4,5\}$]{\includegraphics[width = 0.51\textwidth]{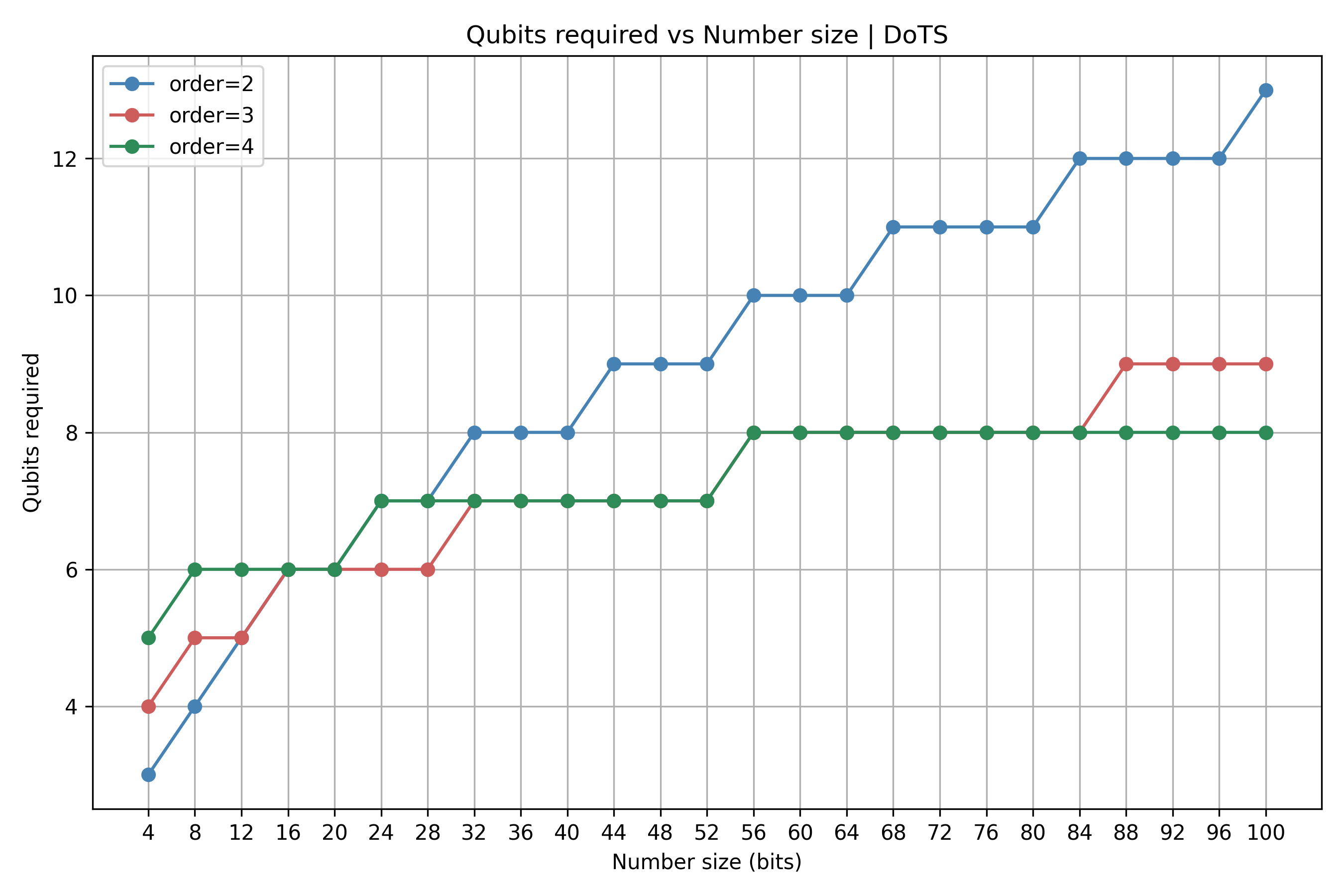}}
    \caption{Scaling of the number of qubits as a function of the number of bits.}
    \label{fig: Fact_DoTS_num_qubits}
\end{figure}

\begin{figure}[H]
    \centering
     \subfloat[Depth for $k\in\{2,3,4,5\}$]{\includegraphics[width = 0.51\textwidth]{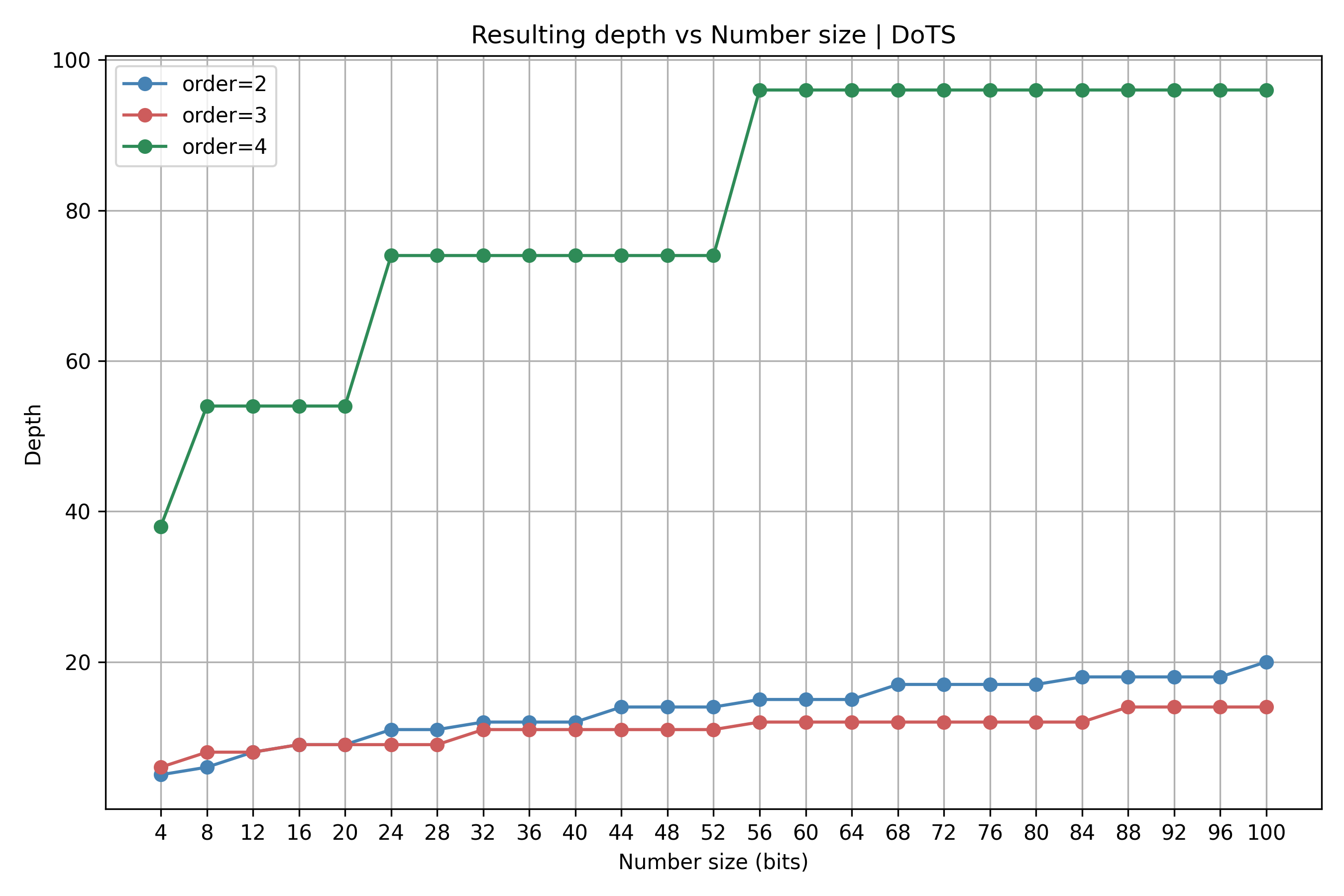}}
     \subfloat[Number of params for $k\in\{2,3,4,5\}$]{\includegraphics[width = 0.51\textwidth]{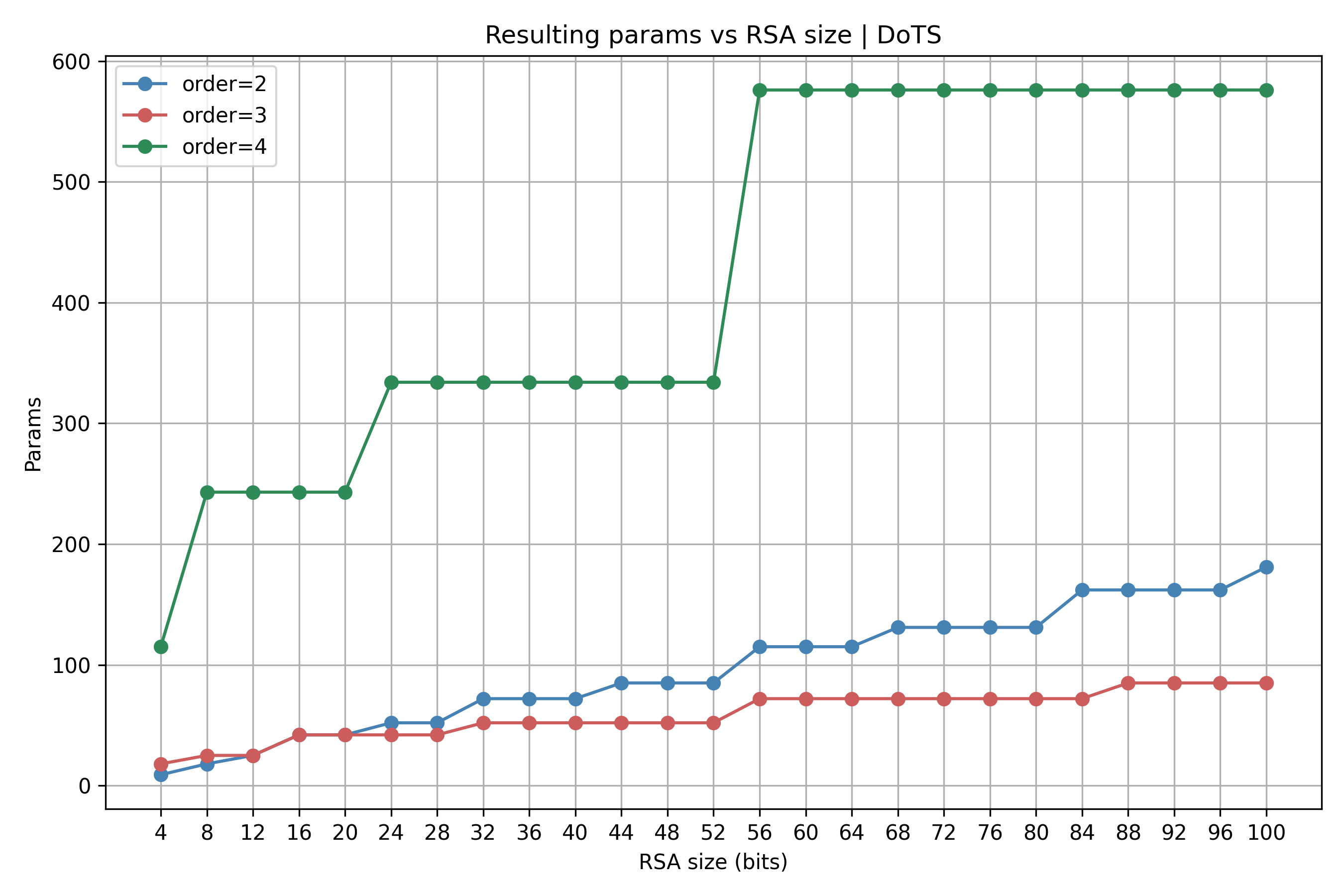}}
    \caption{Scaling of the ideal circuit depth and resultant number of parameters as a function of the number of bits.}
    \label{fig: Fact_DoTS_resources}
\end{figure}

Firstly, the number of qubits, circuit depth (measured as the number of ansatz layers), and variational parameters resulting from the PCE configuration employed in this approach are shown in \textit{Figures} \ref{fig: Fact_DoTS_num_qubits} and \ref{fig: Fact_DoTS_resources}, respectively.

\begin{figure}[H]
    \centering
    \subfloat[$k=2$]{%
        \includegraphics[width=1.0\textwidth]{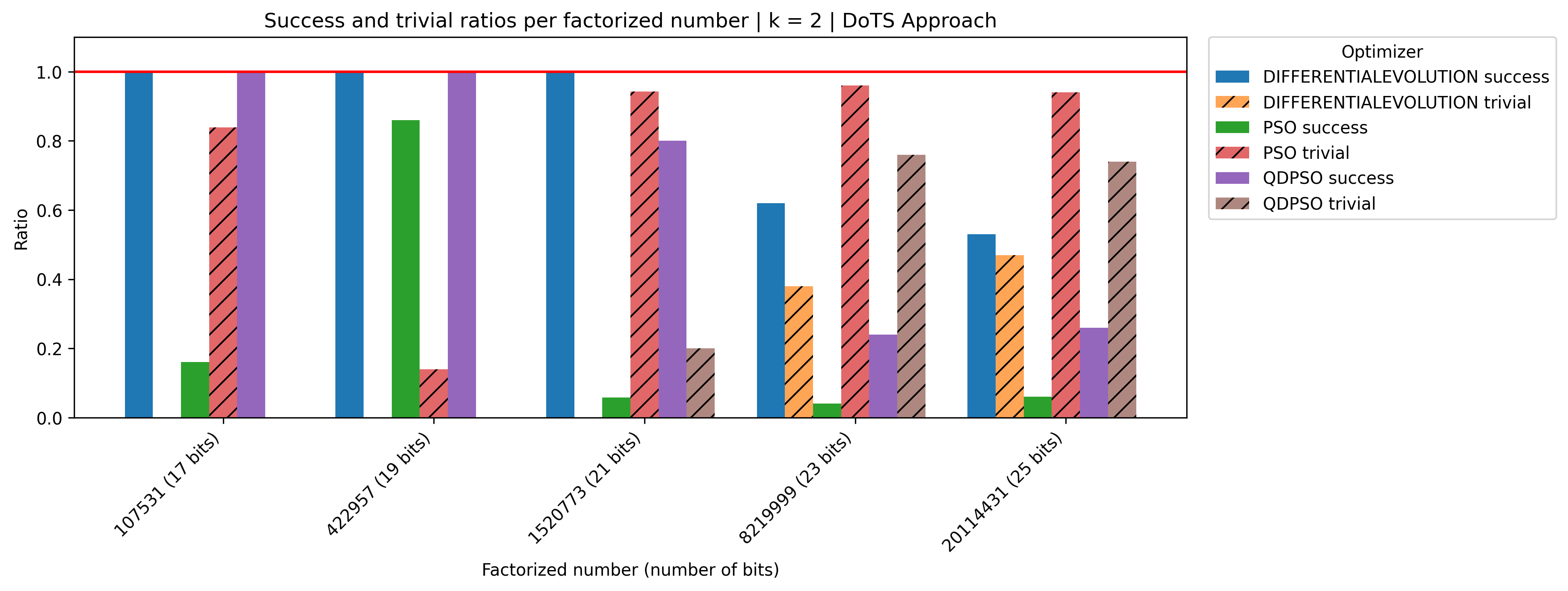}%
    }\\[2ex]
    \subfloat[ $k=3$]{%
        \includegraphics[width=1.0\textwidth]{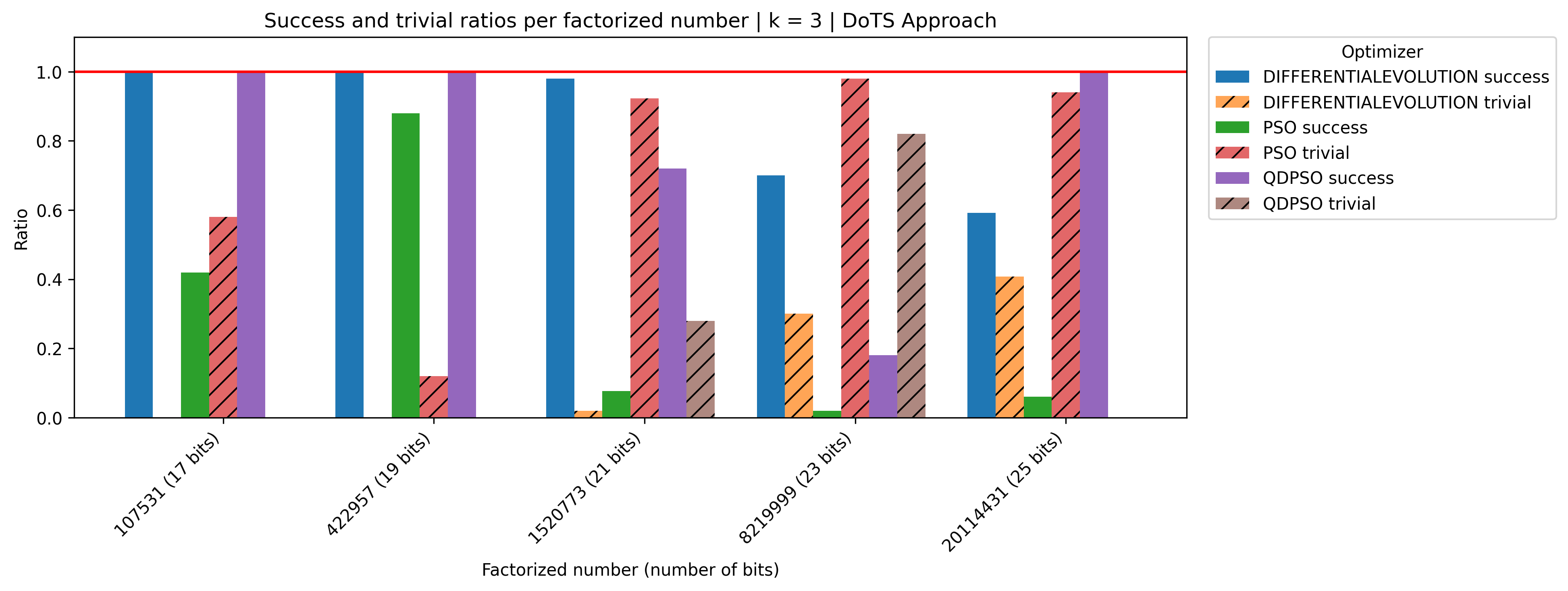}%
    }\\[2ex]
    \subfloat[ $k=4$]{%
        \includegraphics[width=1.0\textwidth]{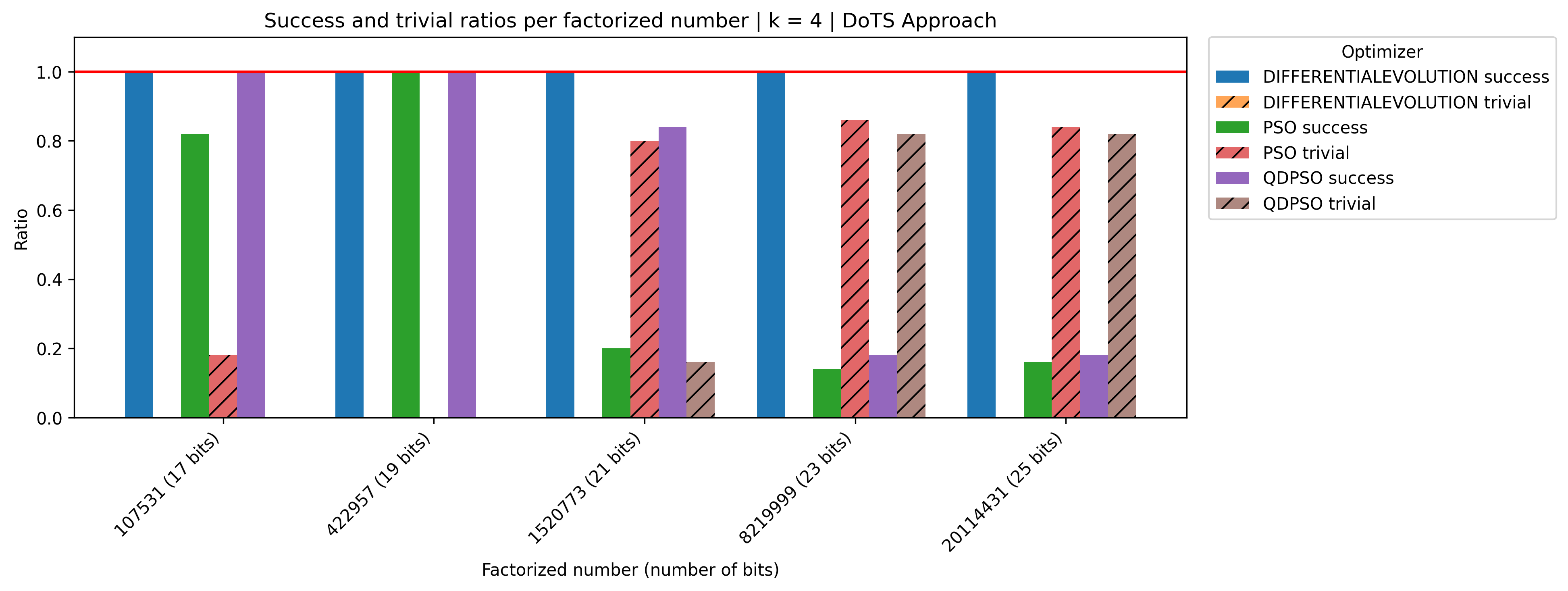}%
    }
    \caption{Factorisation success and trivial solution rates as a function of the number of bits, for different compression orders $k$ and optimisers, over 50 random initialisations.}
    \label{fig:Fact_DoTS_results}
\end{figure}

Using the same semiprime instances considered in Section~\ref{sec_4_1_1}, \textit{Figure}~\ref{fig:Fact_DoTS_results} presents the corresponding results obtained with the DoTS formulation. A substantial performance improvement can be observed across all compression orders and optimisers, with significantly higher factorisation success rates than those obtained with the basic formulation. Although the same qualitative trends remain visible—namely, improved performance for larger compression orders and a gradual decrease in success rate as the bit-length of the semiprime increases—the probability of recovering the correct factorisation is consistently higher. These results suggest that the proposed cost function captures more effectively the arithmetic structure of the problem, leading to a more favourable optimisation landscape. Again, the DE optimiser shows a better performance than the other tested algorithms with the selected configuration of free parameters.

\newpage
~
\vspace{3ex}

 \newpage

\begin{figure}[H]
    \centering
    \includegraphics[width=1.0 \textwidth]{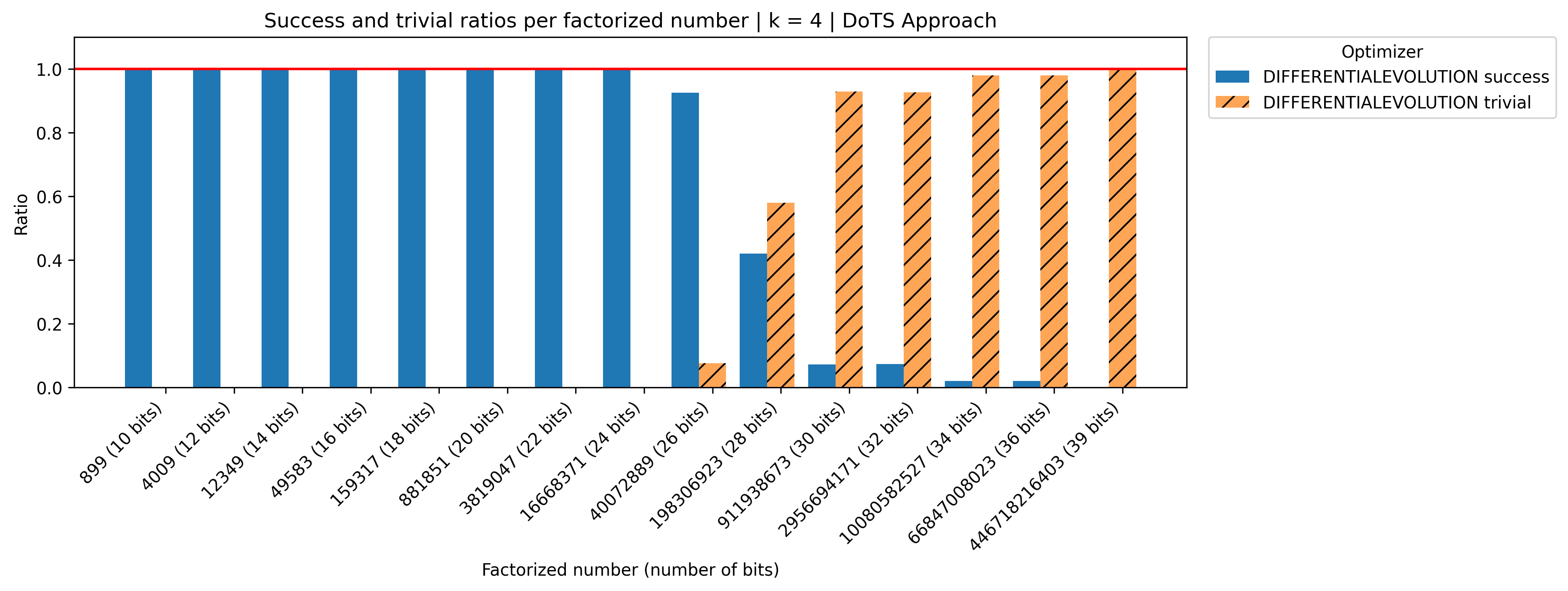}
    \caption{Factorisation success and trivial solution rates as a function of the number of bits, for $k=4$ and DE, over 50 random initialisations.}
    \label{fig:Fact_DoTS_results_DE}
\end{figure}

To assess the scalability of the proposed approach and determine the range of bit-lengths for which successful factorisations can still be achieved, the best-performing configuration, namely $k=4$ combined with the DE optimiser, was selected for further analysis. The semiprime instances considered in this extended study were taken from~\cite{Hong_2025}. The corresponding results are presented in \textit{Figure}~\ref{fig:Fact_DoTS_results_DE}, where the method is capable of successfully factoring integers of up to 36 bits.

One possible concern is that the improved factorisation performance observed for larger compression orders may simply result from a more exhaustive exploration of the search space. Indeed, as commented before, increasing the compression order $k$ leads to a deeper circuit and, consequently, to a larger number of variational parameters. In the DE implementation employed, this increase also leads to larger populations and therefore to a greater number of objective-function evaluations. Together with the mutation and recombination mechanisms inherent to DE, this results in a substantially increased search effort and, consequently, one might argue that the observed solutions arise merely from an increasingly extensive exploration of the possible bit configurations. However, this interpretation is not entirely accurate. Variations in the circuit parameters do not translate directly into bit flips of the binary variables, and therefore the optimisation process cannot be regarded as a purely random exploration of candidate factorisations. 

\begin{figure}[H]
    \centering
    \includegraphics[width=1.0 \textwidth]{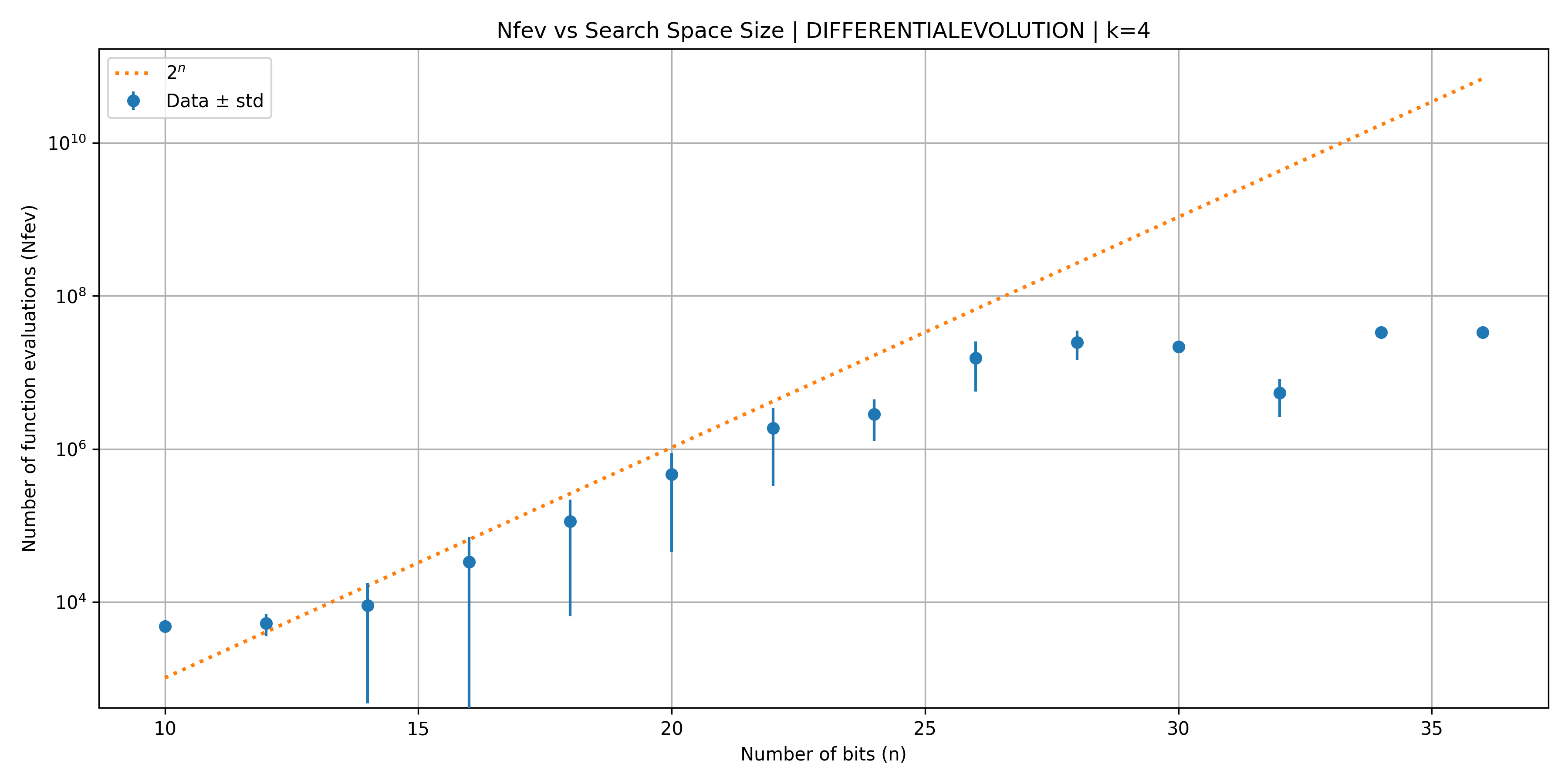}
    \caption{Mean number of objective-function evaluations compared with the number of possible combinations required by a brute-force search, as a function of the bit-length of the semiprime instances, over 50 random initialisations (Logarithmic scale).}
    \label{fig:Fact_escalado}
\end{figure}

This observation is further supported by \textit{Figure}~\ref{fig:Fact_escalado}, which compares the mean number of objective-function evaluations required by the algorithm, together with their standard deviations over 50 random initialisations, against the number of possible combinations, of order $2^n$, that would need to be explored to recover either prime factor through brute-force search. As can be observed, the number of evaluations performed by the algorithm remains several orders of magnitude below the size of the corresponding combinatorial search space. This indicates that the obtained solutions do not arise from an exhaustive exploration of candidate factorisations, but rather from a substantially more efficient guided search process.

\section{Discussion}\label{sec_5}

Previous results permit answering partially the question raised in the first section: Can PCE-based algorithms be used for solving the factorisation problem via optimisation techniques with a limited number of resources (quantum or classical)? 

\begin{figure}[H]
    \centering
     \subfloat[$k\in\{1,2,3,4,5\}$]{\includegraphics[width = 0.51\textwidth]{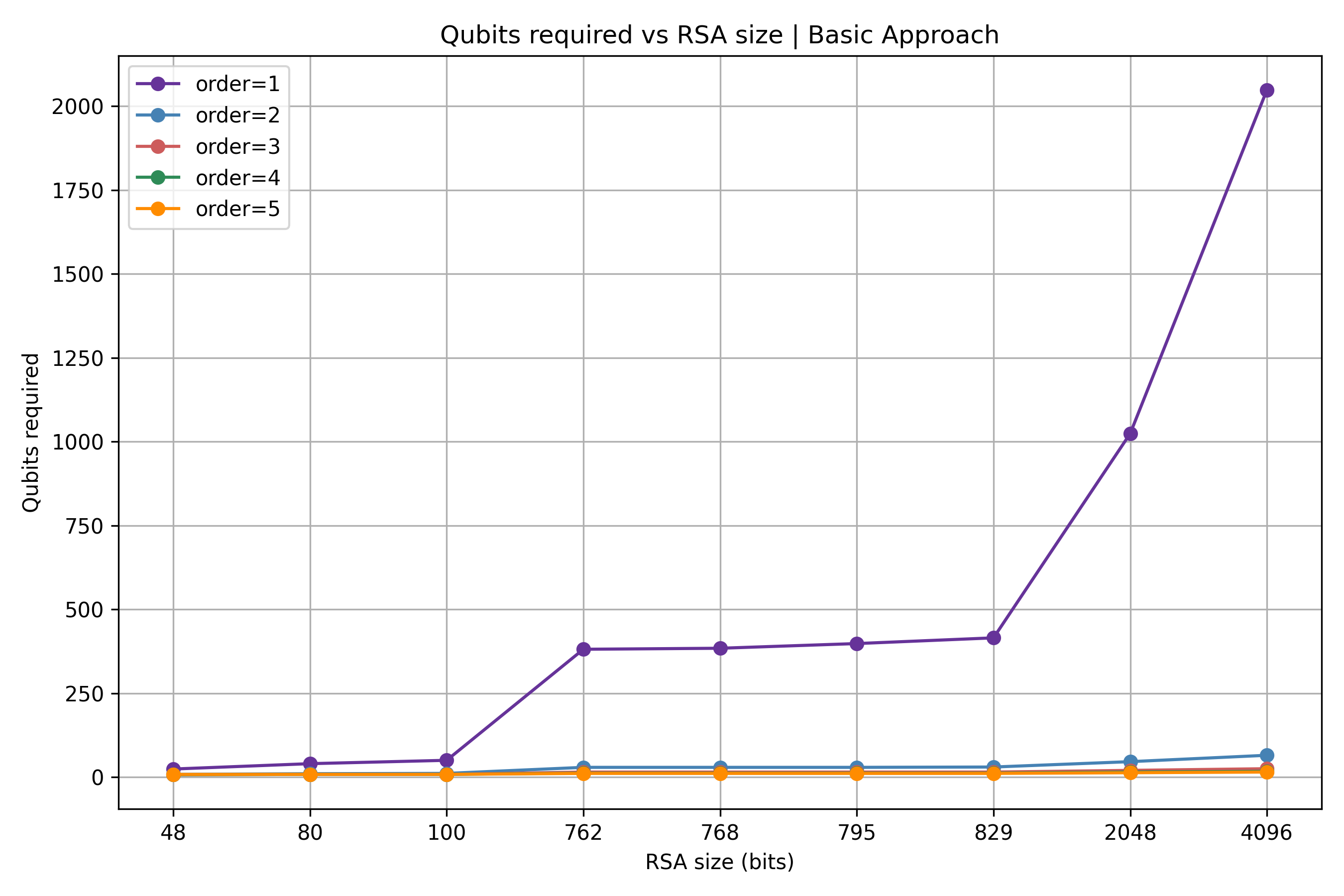}}
     \subfloat[$k\in\{2,3,4,5\}$]{\includegraphics[width = 0.51\textwidth]{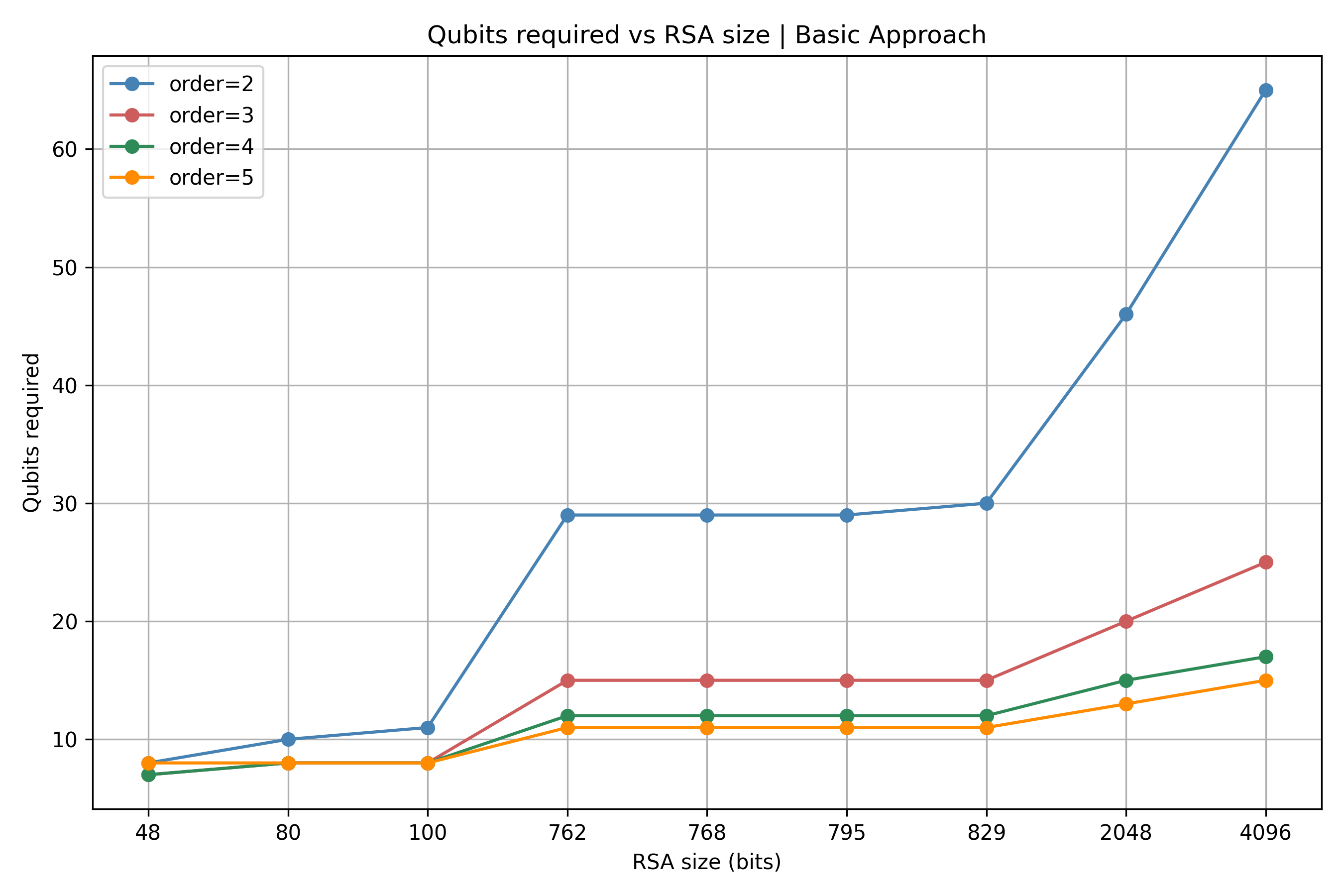}}
    \caption{Scaling of the number of qubits as a function of the number of bits, for the Basic Approach.}
    \label{fig: Fact_org_num_qubits_challenges}
\end{figure}

\begin{figure}[H]
    \centering
     \subfloat[Depth for $k\in\{2,3,4,5\}$]{\includegraphics[width = 0.51\textwidth]{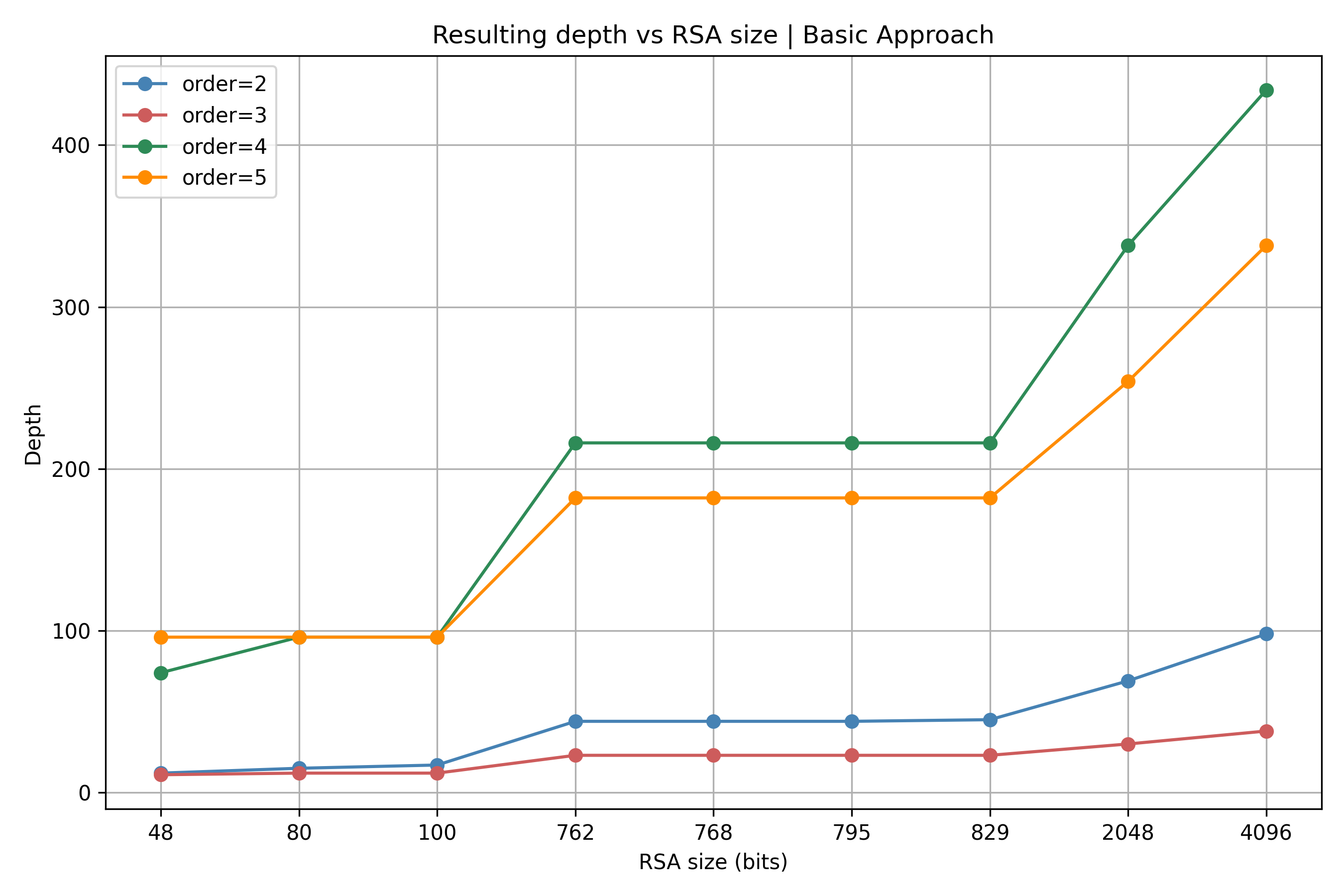}}
     \subfloat[Number of params for $k\in\{2,3,4,5\}$]{\includegraphics[width = 0.51\textwidth]{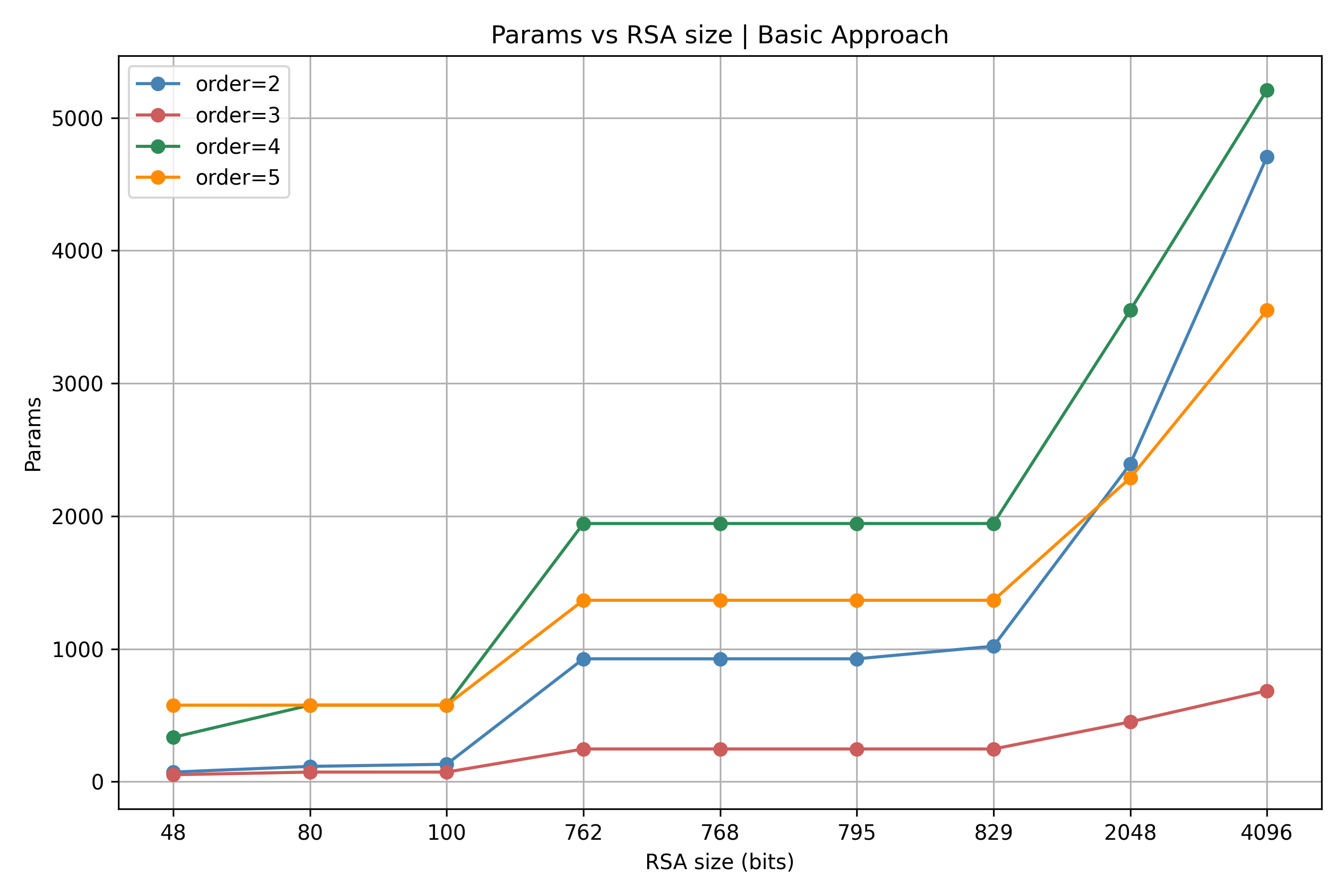}}
    \caption{Scaling of the ideal circuit depth and resultant number of parameters as a function of the number of bits, for the Basic Approach.}
    \label{fig: Fact_org_resources_challenges}
\end{figure}

\begin{figure}[H]
    \centering
     \subfloat[$k\in\{1, 2,3,4,5\}$]{\includegraphics[width = 0.51\textwidth]{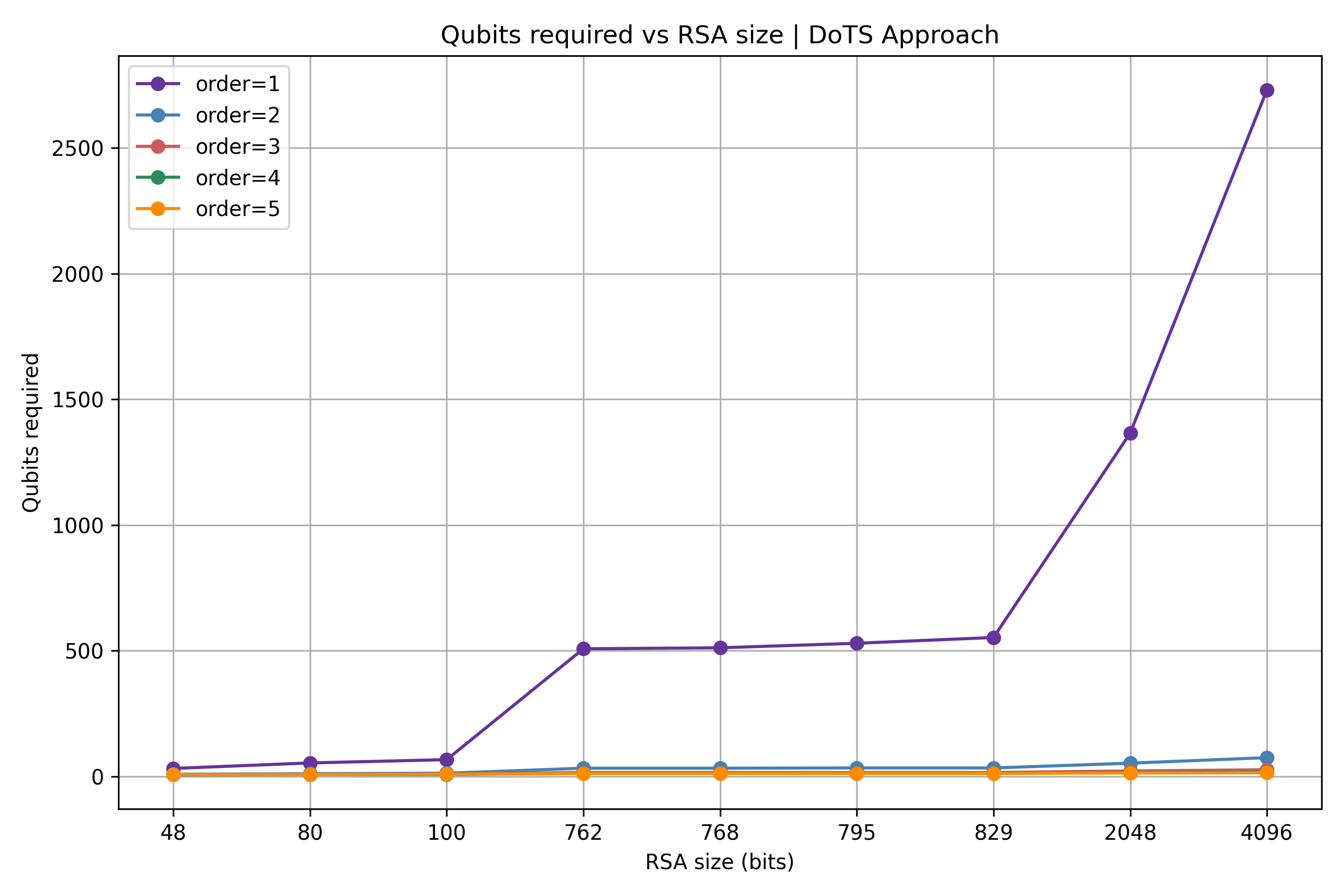}}
     \subfloat[$k\in\{2,3,4,5\}$]{\includegraphics[width = 0.51\textwidth]{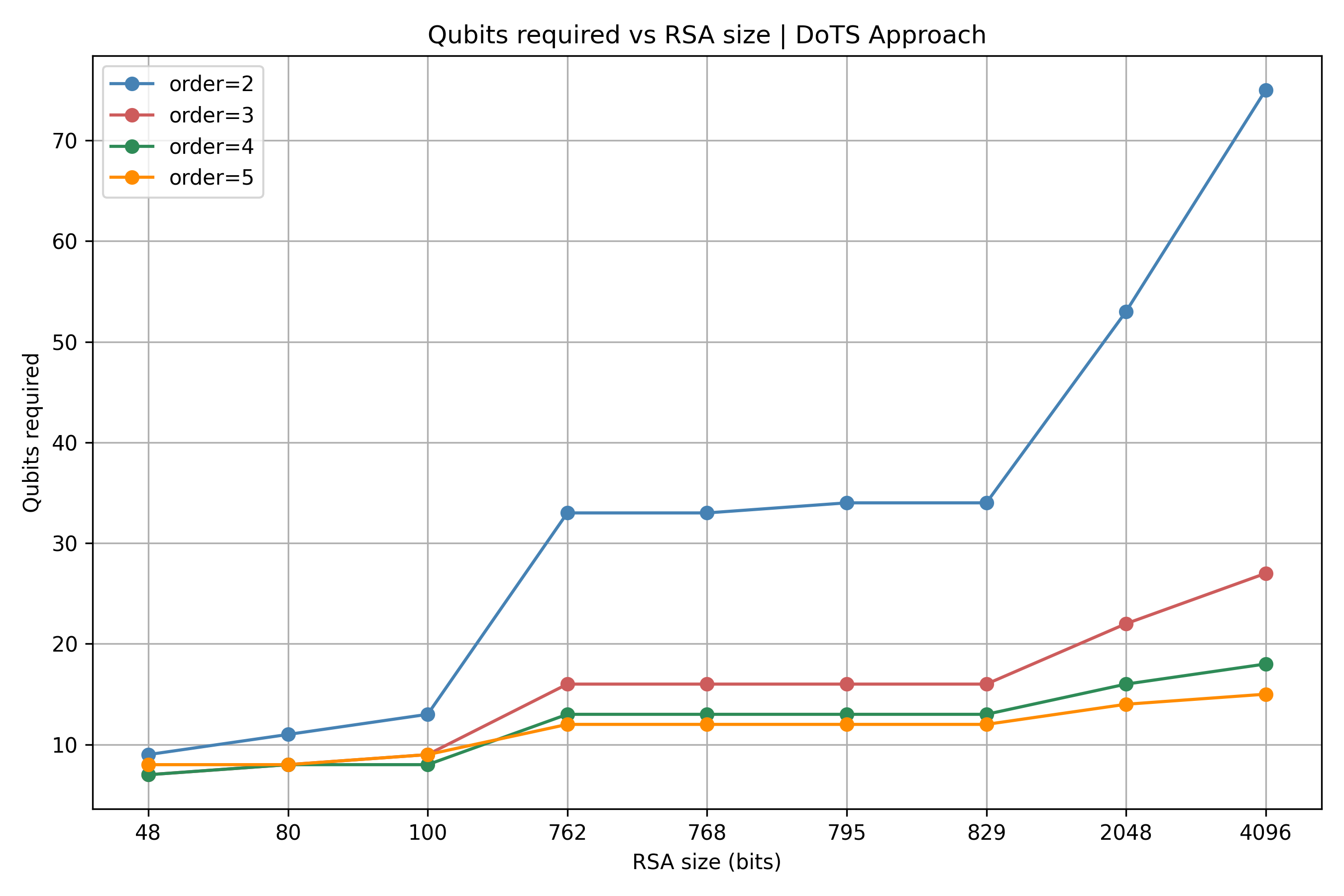}}
    \caption{Scaling of the number of qubits as a function of the number of bits, for the DoTS Approach.}
    \label{fig: Fact_DoTS_num_qubits_challenges}
\end{figure}

\begin{figure}[H]
    \centering
     \subfloat[Depth for $k\in\{2,3,4,5\}$]{\includegraphics[width = 0.51\textwidth]{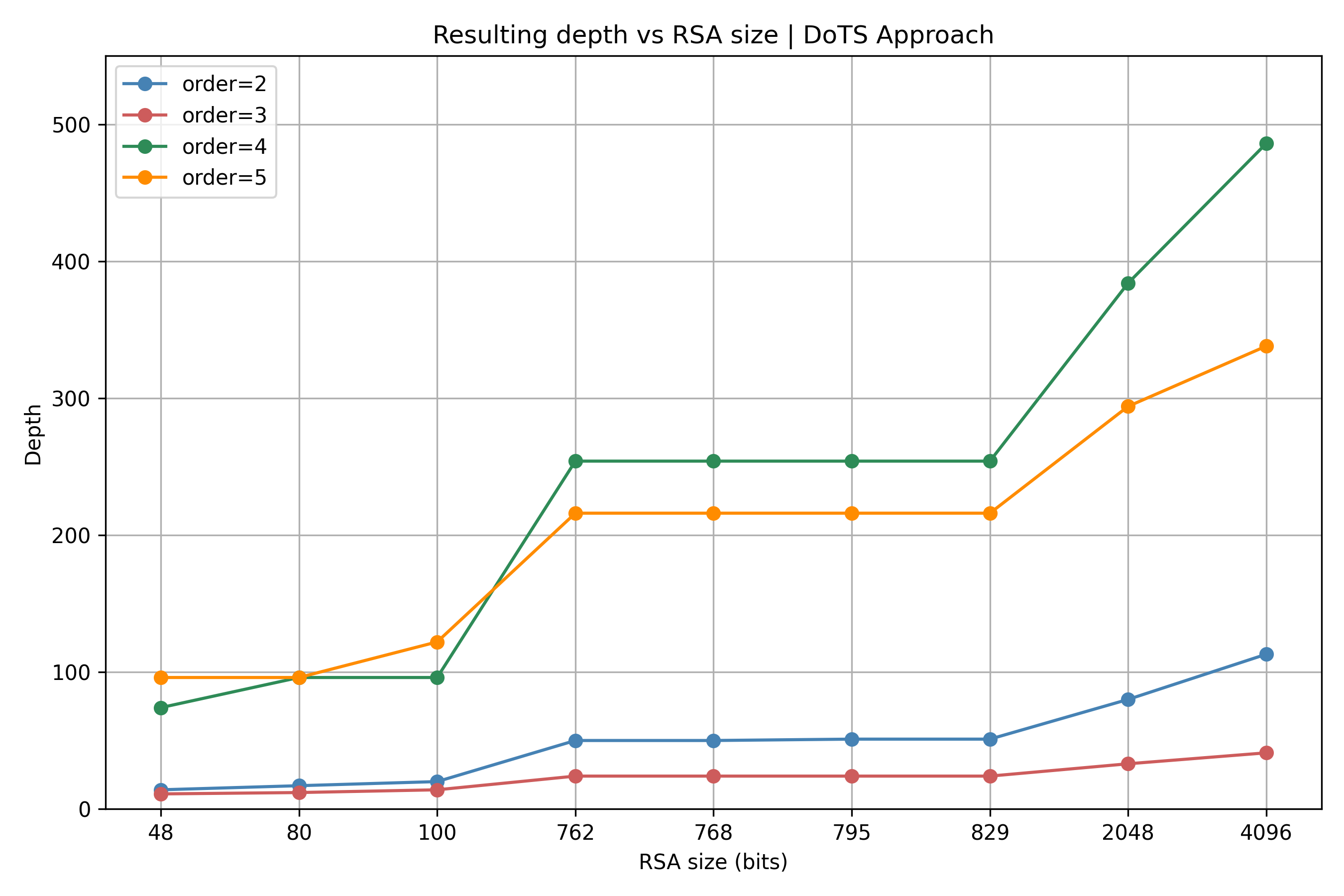}}
     \subfloat[Number of params for $k\in\{2,3,4,5\}$]{\includegraphics[width = 0.51\textwidth]{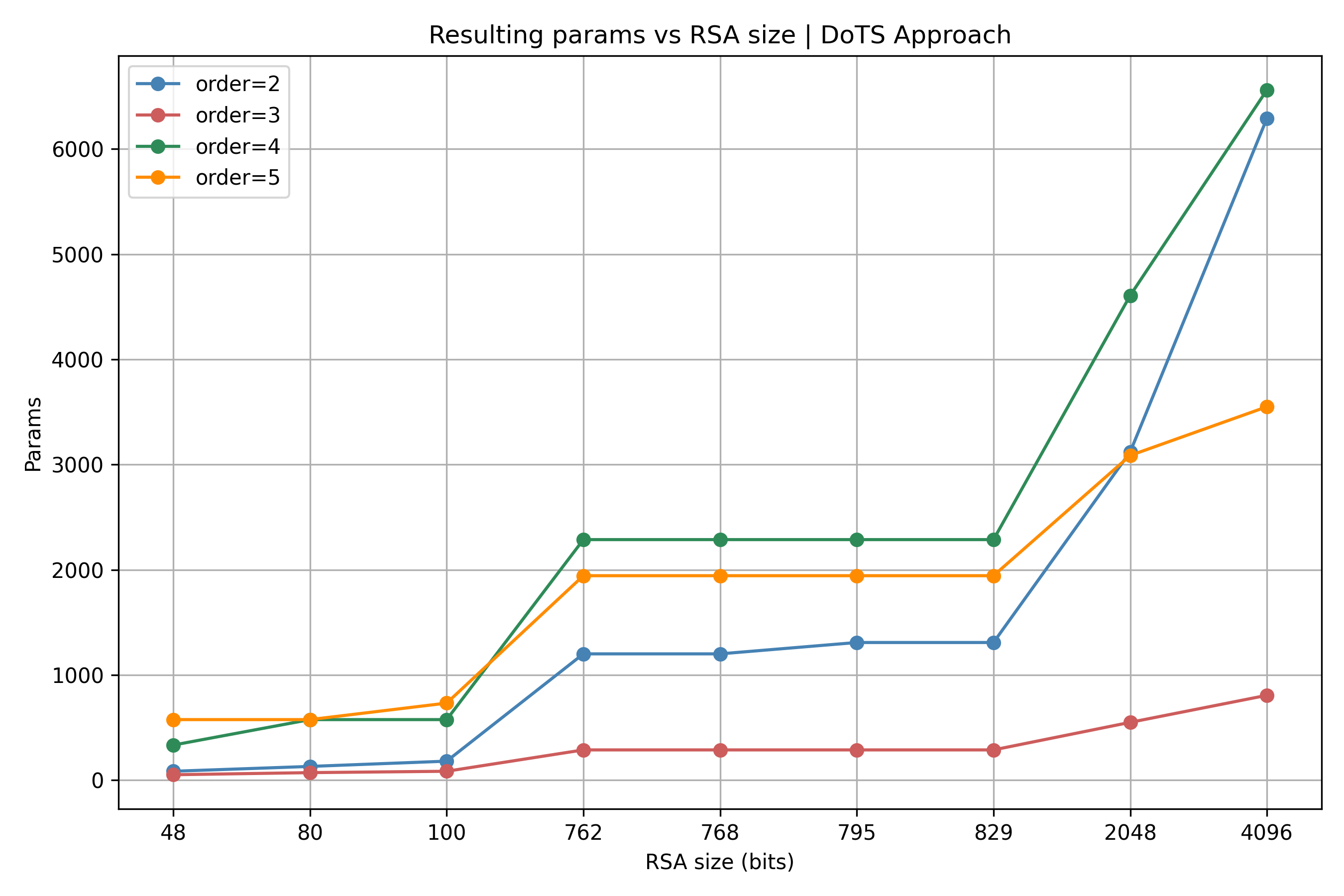}}
    \caption{Scaling of the ideal circuit depth and resultant number of parameters as a function of the number of bits, for the DoTS Approach.}
    \label{fig: Fact_DoTS_resources_challenges}
\end{figure}

We have constructed two approaches using two different classical optimisation cost functions, and the results demonstrate that the selection of the classical function together with the classical optimiser is a key factor for solving the factorisation problem. However, in both cases, the PCE encoding requires only a very limited number of qubits while still providing good solutions even for large compression factors. In fact, Figures \ref{fig: Fact_org_num_qubits_challenges} and \ref{fig: Fact_DoTS_num_qubits_challenges} show that both approaches require an affordable number of qubits to solve their corresponding combinatorial problems for typical RSA key lengths. Furthermore, under the assumed configuration, the required resources — namely, the number of parameters to optimise and the circuit depth — shown in Figures \ref{fig: Fact_org_resources_challenges} and \ref{fig: Fact_DoTS_resources_challenges} remain limited and practically affordable for the analysed problems. Moreover, for compression factors beyond $k=2$, these problems no longer require quantum hardware and can be solved using classical computers. In this case, the search process can be further accelerated by leveraging modern GPU-based emulators, significantly reducing the overall computational cost. Additionally, the number of function evaluations remains limited and does not scale exponentially.

Nevertheless, the two analysed approaches do not provide successful results for the largest tested instances, suggesting that the selected classical formulations may have intrinsic limitations when searching for suitable solutions. This indicates that, in the absence of satisfactory results, more robust alternative formulations should be explored.

Taking these results into account, the answer appears to be affirmative: it is plausible that PCE could be used to address factorisation-as-optimisation problems due to the limited resources required by the quantum(-inspired) stage. When comparing the required quantum resources with alternative approaches such as QAOA or the most general version of VQF, PCE appears to offer a competitive advantage. In fact, the original work \cite{Sciorilli_2025} demonstrates that it is possible to solve optimisation problems involving thousands of binary variables, which is of the same order of magnitude as the number of variables required for factoring large RSA keys. However, even if the quantum component can be handled with currently available resources, the main challenge remains the design of suitable classical cost functions together with effective optimisation strategies. Still, even under a very limited success probability, the possibility of leveraging thousands of GPUs to massively parallelise quantum circuit emulation and perform repeated search attempts could represent a potential risk that should be considered when evaluating the long-term robustness of RSA cryptography. In fact, we have executed the most general factorisation problem. In case of RSA, the problem can be reformulated because the factors usually have less number of bits, usually each one with half of the size of the number to factorise. In fact, with this restriction the trivial solution is unfeasible, and the algorithm maybe could produce better results. We are researching now this modification, but we do not have results yet. 

Additionally, beyond these conclusions, this work has empirically demonstrated that for problems involving a limited number of variables, a PCE-based algorithm combined with the DE optimiser is capable of finding the unique solution of a combinatorial problem with high probability and using only a limited number of function evaluations. More importantly, considering the potential security implications if techniques based on PCE could eventually improve our ability to tackle computational problems underlying RSA-like cryptographic systems, further research on the application of PCE appears well justified.

At the same time, the present study only explores a limited subset of the possible design choices within the PCE framework. Several aspects remain largely unexplored, including alternative problem formulations, more informative cost functions, different optimisation strategies, and alternative circuit architectures, all of which could substantially influence the overall performance of the method. This suggests that the results presented here should be interpreted as an initial proof of concept rather than as a definitive assessment of the capabilities of PCE for large-scale factorisation problems.

\section{Future work}\label{sec_6}

In light of the discussion presented in the previous section, we now analyse several possible directions for future research.

Firstly, regarding the two approaches presented, several possible variants could be explored. The first concerns the \textit{Basic approach}, where we encode the candidate factors using the most general formulation of the factorisation problem. However, in cryptographic applications such as RSA, the prime factors are typically chosen to be balanced, that is, to have approximately the same bit length~\cite{Hinek_2009}. Under this assumption, one has
\begin{equation*}
    l_p \approx l_q \approx \left\lceil \frac{l_N}{2} \right\rceil.
\end{equation*}
Consequently, the number of optimisation variables reduces to approximately $m \approx l_N$, representing a substantial reduction with respect to the general formulation. Such a restriction reduces the size of the search space and may therefore simplify the optimisation process. Secondly, in the \textit{DoTS approach}, it is worth noting that the pair $(a,b)$ satisfying the desired relations is not unique. In fact, if a pair $(a,b)$ satisfies
\begin{equation*}
a^2 \equiv b^2 \pmod N,
\end{equation*}
then any pair congruent to $(a,b)$ modulo $N$ also satisfies the same relation. Indeed, for arbitrary integers $k,\ell\in\mathbb{Z}$,
\begin{equation*}
(a+kN)^2 \equiv a^2 \pmod N,
\qquad
(b+\ell N)^2 \equiv b^2 \pmod N,
\end{equation*}
and therefore
\begin{equation*}
(a+kN)^2 \equiv (b+\ell N)^2 \pmod N.
\end{equation*}
As a consequence, the optimisation landscape contains multiple global solutions associated with the same factorisation. Such degeneracy could be advantageous from an optimisation perspective, as it enlarges the set of configurations capable of yielding the desired factors.

Beyond the specific approaches proposed in this work, a first and perhaps most important direction for future research concerns the design of more informative cost functions, since the results obtained here indicate that the performance of the PCE framework is strongly influenced by the extent to which the cost function captures the arithmetic structure of the factorisation problem. Alternative formulations, such as the original one proposed in the literature, which introduces additional binary variables associated with carry terms, may yield further gains in performance and scalability. More broadly, alternative strategies should also be considered. For instance, approaches based on Schnorr’s algorithm rely on underlying optimisation problems whose structure could potentially be addressed using PCE with a very limited number of qubits, representing another promising direction for future research.

Moreover, the present study has been restricted to the hardware-efficient ansatz originally proposed for PCE and to three gradient-free optimisation algorithms that have previously demonstrated robustness in challenging optimisation landscapes. While these choices provide a reliable baseline for assessing the proposed formulations, they do not exhaust the range of possible circuit architectures and optimisation strategies that could be employed. In particular, recent studies have shown that more advanced variants of DE can outperform standard implementations in variational quantum optimisation tasks~\cite{Novak_2025}, highlighting the potential benefits of adaptive evolutionary strategies. Likewise, hybrid global optimisation frameworks such as \textit{self-adaptive Cooperative Enhanced Scatter Search} (saCeSS)~\cite{Penas_2017} may constitute promising alternatives for navigating the highly non-convex optimisation landscapes encountered in this work. Exploring such approaches, together with alternative ansatz constructions, represents a natural direction for future research.

After defining an ansatz and a competitive optimiser, a more fundamental direction for future research concerns understanding the relationship between compression, circuit complexity, and optimisation performance within the PCE framework. In the present work, the resource analysis has been restricted to the hardware-efficient ansatz and depth scaling originally proposed in the literature, and there is no guarantee that this particular choice is optimal. A systematic characterisation of how circuit depth, variational complexity, and expressive capacity scale with the compression order would therefore be highly valuable. More specifically, an important open question concerns determining how the minimum number of variational parameters required to reliably recover a solution scales with the size of the problem. Such an analysis could provide deeper insight into the fundamental trade-offs between compression, expressibility, and optimisation efficiency, ultimately contributing to the design of more scalable PCE-based algorithms.

Another promising direction concerns the investigation of warm-start strategies, in which the optimisation process is initialised using parameters obtained from previous runs rather than from random configurations. Since RSA keys are typically constructed from balanced prime factors of similar bit length, problem instances of the same size may exhibit common structural features in their optimisation landscapes. Consequently, the parameters associated with a successful factorisation of one semiprime could potentially provide a favourable starting point for the factorisation of other semiprimes of comparable size. Exploring the extent to which such parameter transferability exists could help reduce optimisation times and improve convergence rates. Additionally, preprocessing and post-processing techniques could be incorporated to further improve the overall framework. On one hand, preprocessing methods may help reduce the effective problem size before optimisation, while post-processing strategies could improve the quality of candidate solutions, ultimately increasing the probability of recovering the correct factorisation.

Finally, beyond the algorithmic improvements discussed above, an equally important direction for future research concerns evaluating the behaviour of the proposed method under realistic quantum hardware conditions. In particular, investigating the effect of realistic noise models on the optimisation process would be of significant interest. While excessive noise is generally expected to degrade performance, a moderate level of noise could potentially prove beneficial by introducing perturbations that help the optimiser escape local minima and explore the search space more effectively.


\section*{Acknowledgements}
This work has been mainly financed by  Spanish Ministry for Digital Transformation and of Civil Service of the Spanish Government through the QUANTUM ENIA project call - QuantumSpain, EU through the Recovery, Transformation and Resilience Plan – NextGenerationEU within the framework of the Digital Spain 2026. Also, Andrés Gómez acknowledges the grant PID2024-159713OB-I00 funded by MICIU/AEI/10.13039/501100011033 and by ERDF/EU. 

Additionally, this research project was made possible through the access granted by the Galicia Supercomputing Center (CESGA) to two key parts of its infrastructure. Firstly, its Qmio quantum computing infrastructure with funding from the European Union, through the Operational Programme Galicia 2014-2020 of ERDF\_REACT EU, as part of the European Union's response to the COVID-19 pandemic. Secondly, the supercomputer FinisTerrae III and its permanent data storage system, which have been funded by the NextGeneration EU 2021 Recovery, Transformation and Resilience Plan, ICT2021-006904, and also from the Pluriregional Operational Programme of Spain 2014-2020 of the European Regional Development Fund (ERDF), ICTS-2019-02-CESGA-3, and from the State Programme for the Promotion of Scientific and Technical Research of Excellence of the State Plan for Scientific and Technical Research and Innovation 2013-2016 State subprogramme for scientific and technical infrastructures and equipment of ERDF, CESG15-DE-3114.

\section*{AI Usage}
The original text was written in Spanish and translated by Claude AI to British English. The authors reviewed the generated text in English.
\section*{Conflict of interest}
The authors declare no conflict of interest.

\section*{Availability of data and code}
The data and code can be downloaded from 
\href{https://github.com/csampron/CESGA-Quantum-Spain-PCE-Factorisation}{CESGA-Quantum-Spain-PCE-Factorisation}.


\printbibliography

@article{Sciorilli_2025,
   title={Towards large-scale quantum optimization solvers with few qubits},
   volume={16},
   number={1},
   journal={Nature Communications},
   publisher={Springer Science and Business Media LLC},
   author={Sciorilli, Marco and Borges, Lucas and Patti, Taylor L. and García-Martín, Diego and Camilo, Giancarlo and Anandkumar, Anima and Aolita, Leandro},
   year={2025},
   month=jan }

@article{Failde2023,
  author    = {Faílde, Daniel and Viqueira, José Daniel and Mussa Juane, Mariamo and Gómez, Andrés},
  title     = {Using Differential Evolution to avoid local minima in Variational Quantum Algorithms},
  journal   = {Scientific Reports},
  year      = {2023},
  volume    = {13},
  number    = {1},
  pages     = {16230},
  doi       = {10.1038/s41598-023-43404-3},
  }

@misc{Novak_2025,
      title={Optimization Strategies for Variational Quantum Algorithms in Noisy Landscapes}, 
      author={Vojtěch Novák and Ivan Zelinka and Václav Snášel},
      year={2025},
      eprint={2506.01715},
      archivePrefix={arXiv},
}

@article{Schuld_2021,
   title={Effect of Data Encoding on the Expressive Power of Variational Quantum-Machine-learning Models},
   volume={103},
   number={3},
   journal={Physical Review A},
   publisher={American Physical Society (APS)},
   author={Schuld, Maria and Sweke, Ryan and Meyer, Johannes Jakob},
   year={2021}
}

@article{Larocca_2025,
   title={Barren plateaus in variational quantum computing},
   volume={7},
   DOI={10.1038/s42254-025-00813-9},
   number={4},
   journal={Nature Reviews Physics},
   publisher={Springer Science and Business Media LLC},
   author={Larocca, Martín and Thanasilp, Supanut and Wang, Samson and Sharma, Kunal and Biamonte, Jacob and Coles, Patrick J. and Cincio, Lukasz and McClean, Jarrod R. and Holmes, Zoë and Cerezo, M.},
   year={2025},
   month=mar, pages={174–189} }

@article{Preskill_2018,
   title={Quantum Computing in the NISQ era and beyond},
   volume={2},
   DOI={10.22331/q-2018-08-06-79},
   journal={Quantum},
   publisher={Verein zur Forderung des Open Access Publizierens in den Quantenwissenschaften},
   author={Preskill, John},
   year={2018},
   month=aug, pages={79} }

@article{Bharti_2022,
   title={Noisy intermediate-scale quantum algorithms},
   volume={94},
   DOI={10.1103/revmodphys.94.015004},
   number={1},
   journal={Reviews of Modern Physics},
   publisher={American Physical Society (APS)},
   author={Bharti, Kishor and Cervera-Lierta, Alba and Kyaw, Thi Ha and Haug, Tobias and Alperin-Lea, Sumner and Anand, Abhinav and Degroote, Matthias and Heimonen, Hermanni and Kottmann, Jakob S. and Menke, Tim and Mok, Wai-Keong and Sim, Sukin and Kwek, Leong-Chuan and Aspuru-Guzik, Alán},
   year={2022},
   month=feb }

@article{Cerezo_2021,
   title={Variational quantum algorithms},
   volume={3},
   DOI={10.1038/s42254-021-00348-9},
   number={9},
   journal={Nature Reviews Physics},
   publisher={Springer Science and Business Media LLC},
   author={Cerezo, M. and Arrasmith, Andrew and Babbush, Ryan and Benjamin, Simon C. and Endo, Suguru and Fujii, Keisuke and McClean, Jarrod R. and Mitarai, Kosuke and Yuan, Xiao and Cincio, Lukasz and Coles, Patrick J.},
   year={2021},
   month=aug, pages={625–644} }

@article{Stilck_2021,
   title={Limitations of optimization algorithms on noisy quantum devices},
   volume={17},
   DOI={10.1038/s41567-021-01356-3},
   number={11},
   journal={Nature Physics},
   publisher={Springer Science and Business Media LLC},
   author={Stilck França, Daniel and García-Patrón, Raul},
   year={2021},
   month=oct, pages={1221–1227} }

@misc{Bornens_2023,
      title={Variational quantum algorithms on cat qubits}, 
      author={Anne-Solène Bornens and Michel Nowak},
      year={2023},
      eprint={2305.14143},
      archivePrefix={arXiv},
}

@misc{Zhang_2022,
      title={Fundamental limitations on optimization in variational quantum algorithms}, 
      author={Hao-Kai Zhang and Chengkai Zhu and Geng Liu and Xin Wang},
      year={2022},
      eprint={2205.05056},
      archivePrefix={arXiv},
}

@techreport{Burges_2002,
author = {Burges, Chris J.C.},
title = {Factoring as Optimization},
year = {2002},
month = {January},
pages = {19},
number = {MSR-TR-2002-83},
}

@misc{soloviev2025,
      title={Large-scale portfolio optimization using Pauli Correlation Encoding}, 
      author={Vicente P. Soloviev and Michal Krompiec},
      year={2025},
      eprint={2511.21305},
      archivePrefix={arXiv},
      primaryClass={quant-ph},
      url={https://arxiv.org/abs/2511.21305}, 
}

@misc{Padin_2026,
      title={Pauli Correlation Encoding for Budget-Constrained Optimization}, 
      author={Jacobo Padín-Martínez and Vicente P. Soloviev and Alejandro Borrallo-Rentero and Antón Rodríguez-Otero and Raquel Alfonso-Rodríguez and Michal Krompiec},
      year={2026},
      eprint={2602.17479},
      archivePrefix={arXiv}, 
}

@misc{Gidney_2025,
      title={How to factor 2048 bit RSA integers with less than a million noisy qubits}, 
      author={Craig Gidney},
      year={2025},
      eprint={2505.15917},
      archivePrefix={arXiv},
      primaryClass={quant-ph}
}

@misc{Bansimba_2025,
      title={Integer Factorization: Another perspective}, 
      author={Gilda Rech Bansimba and Regis Freguin Babindamana},
      year={2025},
      eprint={2507.07055},
      archivePrefix={arXiv}
}

@misc{Gupta_2009,
      title={Revisiting Fermat's Factorization for the RSA Modulus}, 
      author={Sounak Gupta and Goutam Paul},
      year={2009},
      eprint={0910.4179},
      archivePrefix={arXiv}
}

@article{Alberts_2026,
   title={Explicit analytic continuation of Euler products},
   volume={5},
   DOI={10.2140/ent.2026.5.49},
   number={1},
   journal={Essential Number Theory},
   publisher={Mathematical Sciences Publishers},
   author={Alberts, Brandon},
   year={2026},
   month=Mar, pages={49–112} }

@inproceedings{Pomerance_1998,
  title={A Tale of Two Sieves},
  author={Carl Pomerance and Paul Erd{\"o}s},
  year={1998},
}

@Inbook{Lenstra_2005,
author="Lenstra, Arjen K.",
editor="van Tilborg, Henk C. A.",
title="Integer Factoring",
bookTitle="Encyclopedia of Cryptography and Security",
year="2005",
publisher="Springer US",
address="Boston, MA",
pages="290--297",
doi="10.1007/0-387-23483-7_200",
}

@inproceedings{Pomerance_1984,
author = {Pomerance, Carl},
year = {1984},
month = {01},
pages = {169-182},
title = {The Quadratic Sieve Factoring Algorithm.}
}

@article{Lenstra_2001,
author = {Lenstra, Arjen and Lenstra, Hendrik and Manasse, Mark and Pollard, J.},
year = {2001},
month = {08},
title = {The Number Field Sieve}
}

@article{Peruzzo_2014,
  author    = {Alberto Peruzzo and Jarrod McClean and Peter Shadbolt and
               Man-Hong Yung and Xiao-Qi Zhou and Peter J. Love and
               Al{\'a}n Aspuru-Guzik and Jeremy L. O'Brien},
  title     = {A Variational Eigenvalue Solver on a Photonic Quantum Processor},
  journal   = {Nature Communications},
  volume    = {5},
  pages     = {4213},
  year      = {2014},
  doi       = {10.1038/ncomms5213}
}

@article{Anschuetz_2018,
  author    = {Eric R. Anschuetz and Jonathan P. Olson and
               Al{\'a}n Aspuru-Guzik and Yudong Cao},
  title     = {Variational Quantum Factoring},
  journal   = {Quantum Science and Technology},
  volume    = {5},
  number    = {3},
  pages     = {034012},
  year      = {2020},
  doi       = {10.48550/arXiv.1808.08927}
}

@article{Anschuetz_2022,
author = {Anschuetz, Eric and Kiani, Bobak},
year = {2022},
month = {12},
pages = {7760},
title = {Quantum variational algorithms are swamped with traps},
volume = {13},
journal = {Nature Communications},
doi = {10.1038/s41467-022-35364-5}
}

@article{Weidenfeller_2022,
   title={Scaling of the quantum approximate optimization algorithm on superconducting qubit based hardware},
   volume={6},
   DOI={10.22331/q-2022-12-07-870},
   journal={Quantum},
   publisher={Verein zur Forderung des Open Access Publizierens in den Quantenwissenschaften},
   author={Weidenfeller, Johannes and Valor, Lucia C. and Gacon, Julien and Tornow, Caroline and Bello, Luciano and Woerner, Stefan and Egger, Daniel J.},
   year={2022},
   month=Dec, pages={870} }

@article{Sengupta_2018,
   title={Particle Swarm Optimization: A Survey of Historical and Recent Developments with Hybridization Perspectives},
   volume={1},
   DOI={10.3390/make1010010},
   number={1},
   journal={Machine Learning and Knowledge Extraction},
   publisher={MDPI AG},
   author={Sengupta, Saptarshi and Basak, Sanchita and Peters, Richard},
   year={2018},
   month=Oct, pages={157–191} }

@phdthesis{Jones_2024,
author = {Jones, Andrew},
year = {2023},
month = {12},
pages = {},
title = {Particle Swarm Optimization},
doi = {10.13140/RG.2.2.32162.20163}
}

@misc{sotostzam_pso,
  author = {Sotostzam},
  title  = {Particle Swarm Optimization},
  year   = {2026},
  url    = {https://github.com/sotostzam/particle-swarm-optimization},
  note   = {GitHub repository}
}

@article{Yang_2004,
author = {Yang, S.-Y and Jiao, L.-C},
year = {2004},
month = {10},
pages = {201-204},
title = {Quantum particle swarm optimization algorithm and its application},
volume = {32}
}

@INPROCEEDINGS{Sun_2004,
  author={Jun Sun and Bin Feng and Wenbo Xu},
  booktitle={Proceedings of the 2004 Congress on Evolutionary Computation (IEEE Cat. No.04TH8753)}, 
  title={Particle swarm optimization with particles having quantum behavior}, 
  year={2004},
  volume={1},
  pages={325-331 Vol.1},
  doi={10.1109/CEC.2004.1330875}}

@misc{ngroup_qpso,
  author = {Chun Nien},
  title  = {QPSO: A Python Implementation of Quantum Particle Swarm Optimization},
  year   = {2017},
  url    = {https://github.com/ngroup/qpso},
  note   = {GitHub repository}
}

@article{Shor_1997,
   title={Polynomial-Time Algorithms for Prime Factorization and Discrete Logarithms on a Quantum Computer},
   volume={26},
   DOI={10.1137/s0097539795293172},
   number={5},
   journal={SIAM Journal on Computing},
   publisher={Society for Industrial & Applied Mathematics (SIAM)},
   author={Shor, Peter W.},
   year={1997},
   month=Oct, pages={1484–1509} }

@misc{Bagourd_2025,
      title={Practical Challenges in Executing Shor's Algorithm on Existing Quantum Platforms}, 
      author={Paul Bagourd and Julian Jang-Jaccard and Vincent Lenders and Alain Mermoud and Torsten Hoefler and Cornelius Hempel},
      year={2026},
      eprint={2512.15330},
      archivePrefix={arXiv},
}

@book{Hinek_2009,
  title={Cryptanalysis of RSA and Its Variants},
  author={Hinek, M.J.},
  series={Chapman \& Hall/CRC Cryptography and Network Security Series},
  year={2009},
  publisher={CRC Press}
}

@misc{bigprimes_rsa,
  title        = {RSA Challenge Generator},
  author       = {{BigPrimes.net}},
  year         = {2026},
  url          = {https://bigprimes.org/RSA-challenge}
}

@article{Hong_2025,
author = {Hong, Chunlei and Pei, Zhi and Wang, Qidi and Yang, Shuxiao and Yu, Jingjing and Wang, Chao},
year = {2025},
month = {01},
pages = {},
title = {Quantum attack on RSA by D-Wave Advantage: a first break of 80-bit RSA},
volume = {68},
journal = {Science China Information Sciences},
doi = {10.1007/s11432-024-4163-6}
}

@article{Penas_2017,
author = {Penas, David R. and González, Patricia and Egea, Jose A and Doallo, Ramón and Banga, Julio},
year = {2017},
month = {01},
pages = {52},
title = {Parameter estimation in large-scale systems biology models: A parallel and self-adaptive cooperative strategy},
volume = {18},
journal = {BMC Bioinformatics},
doi = {10.1186/s12859-016-1452-4}
}

@article{Karamlou_2021,
   title={Analyzing the performance of variational quantum factoring on a superconducting quantum processor},
   volume={7},
   DOI={10.1038/s41534-021-00478-z},
   number={1},
   journal={npj Quantum Information},
   publisher={Springer Science and Business Media LLC},
   author={Karamlou, Amir H. and Simon, William A. and Katabarwa, Amara and Scholten, Travis L. and Peropadre, Borja and Cao, Yudong},
   year={2021},
   month=Oct }

@misc{Willsch_2025,
      title={The State of Factoring on Quantum Computers}, 
      author={Dennis Willsch and Philipp Hanussek and Georg Hoever and Madita Willsch and Fengping Jin and Hans De Raedt and Kristel Michielsen},
      year={2025},
      eprint={2410.14397},
      archivePrefix={arXiv},
      primaryClass={quant-ph},
      doi={https://doi.org/10.34734/FZJ-2025-01965},
}

@misc{Yan_2022,
      title={Factoring integers with sublinear resources on a superconducting quantum processor}, 
      author={Bao Yan and Ziqi Tan and Shijie Wei and Haocong Jiang and Weilong Wang and Hong Wang and Lan Luo and Qianheng Duan and Yiting Liu and Wenhao Shi and Yangyang Fei and Xiangdong Meng and Yu Han and Zheng Shan and Jiachen Chen and Xuhao Zhu and Chuanyu Zhang and Feitong Jin and Hekang Li and Chao Song and Zhen Wang and Zhi Ma and H. Wang and Gui-Lu Long},
      year={2022},
      eprint={2212.12372},
      archivePrefix={arXiv},
      primaryClass={quant-ph} 
}

@article{Tesoro_2026,
   title={Integer factorization via tensor-network Schnorr’s sieving},
   volume={113},
   DOI={10.1103/d9dl-ctt4},
   number={3},
   journal={Physical Review A},
   publisher={American Physical Society (APS)},
   author={Tesoro, Marco and Siloi, Ilaria and Jaschke, Daniel and Magnifico, Giuseppe and Montangero, Simone},
   year={2026},
   month=Mar }

@misc{Scholten_2024,
      title={Assessing the Benefits and Risks of Quantum Computers}, 
      author={Travis L. Scholten and Carl J. Williams and Dustin Moody and Michele Mosca and William Hurley and William J. Zeng and Matthias Troyer and Jay M. Gambetta},
      year={2024},
      eprint={2401.16317},
      archivePrefix={arXiv},
      primaryClass={quant-ph}
}

@article{Amico_2019,
   title={Experimental study of Shor’s factoring algorithm using the IBM Q Experience},
   volume={100},
   DOI={10.1103/physreva.100.012305},
   number={1},
   journal={Physical Review A},
   publisher={American Physical Society (APS)},
   author={Amico, Mirko and Saleem, Zain H. and Kumph, Muir},
   year={2019},
   month=July }

@article{Willsch_2023,
   title={Large-Scale Simulation of Shor’s Quantum Factoring Algorithm},
   volume={11},
   DOI={10.3390/math11194222},
   number={19},
   journal={Mathematics},
   publisher={MDPI AG},
   author={Willsch, Dennis and Willsch, Madita and Jin, Fengping and De Raedt, Hans and Michielsen, Kristel},
   year={2023},
   month=Oct, pages={4222} }

@article{Smolin_2013,
   title={Oversimplifying quantum factoring},
   volume={499},
   DOI={10.1038/nature12290},
   number={7457},
   journal={Nature},
   publisher={Springer Science and Business Media LLC},
   author={Smolin, John A. and Smith, Graeme and Vargo, Alexander},
   year={2013},
   month=July, pages={163–165} }

@misc{Sobhani_2025,
      title={Variational Quantum Eigensolver Approach to Prime Factorization on IBM's Noisy Intermediate Scale Quantum Computer}, 
      author={Mona Sobhani and Yahui Chai and Tobias Hartung and Karl Jansen},
      year={2025},
      eprint={2410.01935},
      archivePrefix={arXiv},
      primaryClass={quant-ph}
}

@misc{Phan_2022,
      title={On quantum factoring using noisy intermediate scale quantum computers}, 
      author={Vivian Phan and Arttu Pönni and Matti Raasakka and Ilkka Tittonen},
      year={2022},
      eprint={2208.07085},
      archivePrefix={arXiv},
      primaryClass={quant-ph}
}

@inproceedings{Schnorr_1997,
author = {Schnorr, Claus},
year = {1997},
month = {08},
pages = {},
title = {Factoring Integers and Computing Discrete Logarithms via Diophantine Approximation},
isbn = {978-3-540-54620-7},
doi = {10.1007/3-540-46416-6_24}
}

@misc{Vera_2010,
      title={A Note on Integer Factorization Using Lattices}, 
      author={Antonio Ignacio Vera},
      year={2010},
      eprint={1003.5461},
      archivePrefix={arXiv},
      primaryClass={cs.DS}, 
}

@article{Schnorr_2013,
author = {Schnorr, Claus},
year = {2013},
month = {01},
title = {Average Time Fast SVP and CVP Algorithms: Factoring Integers in Polynomial Time},
journal = {Lecture Notes in Computer Science},
doi = {10.1007/978-3-642-42001-6_6}
}

@misc{Dattani_2014,
      title={Quantum factorization of 56153 with only 4 qubits}, 
      author={Nikesh S. Dattani and Nathaniel Bryans},
      year={2014},
      eprint={1411.6758},
      archivePrefix={arXiv},
      primaryClass={quant-ph}
}

@article{Xu_2012,
   title={Erratum: Quantum Factorization of 143 on a Dipolar-Coupling Nuclear Magnetic Resonance System [Phys. Rev. Lett.<b>108</b>, 130501 (2012)]},
   volume={109},
   DOI={10.1103/physrevlett.109.269902},
   number={26},
   journal={Physical Review Letters},
   publisher={American Physical Society (APS)},
   author={Xu, Nanyang and Zhu, Jing and Lu, Dawei and Zhou, Xianyi and Peng, Xinhua and Du, Jiangfeng},
   year={2012},
   month=Dec }

@article{Hardy_1917,
  title={On the normal number of prime factors of a number n},
  author={Hardy, Godfrey H},
  journal={Quart. J. Math. Oxford Ser.},
  volume={48},
  pages={76--92},
  year={1917}
}

@misc{Alonso_2026,
      title={Benchmark of Pauli Correlation Encoding for different optimisation problems}, 
      author={Fernando Alonso and Colomán Samprón and Jacobo Veiga and Mariamo Mussa Juane and Andrés Gómez},
      year={2026},
      eprint={2606.18914},
      archivePrefix={arXiv},
      primaryClass={quant-ph},
}

\end{document}